\documentclass[letterpaper,11pt,leqno]{article}
\usepackage[numbers, sort&compress]{natbib}
\usepackage{paper}
\pdfoutput=1 
\usepackage[table,xcdraw]{xcolor}
\usepackage[normalem]{ulem}
\usepackage{tabularx}

\hypersetup{pdftitle={Minimalist LaTeX Template for Academic Papers}}

\newcommand{\bib}{paper.bib}

\begin{document}

\title{Dual quantum locking: Dynamic coupling of hydrogen and water sublattices in hydrogen filled ice}

\author{
Loan Renaud\textsuperscript{a,1,2}
\and
Tomasz Poreba\textsuperscript{b,1}
\and
Simone Di Cataldo\textsuperscript{c}
\and
Alasdair Nicholls\textsuperscript{b}
\and
Léon Andriambariarijaona\textsuperscript{d}
\and
Maria Rescigno\textsuperscript{b,c,e}
\and
Richard Gaal\textsuperscript{b}
\and
Michele Casula\textsuperscript{a}
\and
A. Marco Saitta\textsuperscript{a}
\and
Livia Eleonora Bove\textsuperscript{a,b,c,2}
\thanks{\textsuperscript{a}Institut de Min\'{e}ralogie, de Physique des Mat\'{e}riaux et de Cosmochimie (IMPMC), Sorbonne Universit\'{e}, CNRS UMR 7590,  MNHN, 4, place Jussieu, Paris, France
\\
\textsuperscript{b}Laboratory of Quantum Magnetism, Institute of Physics, \'{E}cole Polytechnique F\'{e}d\'{e}erale de Lausanne, CH-1015 Lausanne, Switzerland
\\
\textsuperscript{c}Dipartimento di Fisica, Sapienza Universit\`a di Roma, Piazzale Aldo Moro 5, 00185 Roma, Italy
\\
\textsuperscript{d}Laboratoire pour l'Utilisation des Lasers Intenses (LULI), École Polytechnique, 91120 Palaiseau, France
\\
\textsuperscript{e}Institut Laue-Langevin, 71 Avenue des Martyrs, Cedex 9, Grenoble, France
\\[2ex] 
\textsuperscript{1}L.R. and T.P. contributed equally to this work.
\\
\textsuperscript{2}Corresponding authors. E-mail: \href{mailto:loan.renaud@sorbonne-universite.fr}{loan.renaud@sorbonne-universite.fr}, \href{mailto:livia.bove@sorbonne-universite.fr}{livia.bove@sorbonne-universite.fr}}}

\date{October 2025}   

\begin{titlepage}
\maketitle

\newpage
\section*{Abstract}
 Hydrogen hydrates (HH) are a unique class of materials composed of hydrogen molecules confined within crystalline water frameworks. Among their multiple phases, the filled ice structures, particularly the cubic C2 phase, exhibit exceptionally strong host-guest interactions due to ultra-short H$_2$-H$_2$O distances and a 1:1 stoichiometry leading to two interpenetrated identical diamond-like sublattices, one comprised of water molecules, the other of hydrogen molecules. At high pressures, nuclear quantum effects involving both hydrogen molecules and the water lattice become dominant, giving rise to a dual-lattice quantum system. In this work, we explore the sequence of pressure- and temperature-driven phase transitions in HH, focusing on the interplay between molecular rotation, orientational ordering, lattice symmetry breaking and hydrogen bond symmetrization. Using a combination of computational modeling based on classical and path-integral molecular dynamics,
 quantum embedding, and high pressure experiments, including Raman spectroscopy and synchrotron X-ray diffraction at low temperatures and high pressures, we identify signatures of quantum-induced ordering and structural transformations in the C2 phase. Our findings reveal that orientational ordering in HH occurs at much lower pressures than in solid hydrogen, by inducing structural changes in the water network and enhancing the coupling of water and hydrogen dynamics. This work provides new insights into the quantum behavior of hydrogen under extreme mechanochemical confinement and establishes hydrogen-filled ices as a promising platform for the design of hydrogen-rich quantum materials. 

\end{titlepage}

The rotational dynamics of guest gas molecules confined in porous materials are profoundly influenced by their local environment \cite{Bove2024}. In large, nearly spherical cages, guest species often behave as nearly free rotors \cite{Jameson2004,ranieri_quantum_2019, ranieri_large-cage_2024}. In contrast, low-symmetry or anisotropic environments impose steric constraints that hinder rotation, leading to librational motions or even complete orientational freezing \cite{Ulivi2007, Xu2013, Longo2023}.

Such effects have been extensively studied in clathrate hydrates, where the host–guest potential not only quantizes the center-of-mass motion but can also induce partial orientational ordering at low temperatures and high pressures \cite{Tse_MethaneHydrates2002, hirai2023, Ulivi2007,Strobel2009}. Zeolite frameworks, with their inherent anisotropy and stronger host–guest interactions, exhibit similar behavior: small guest molecules often display restricted rotational freedom and develop orientational preferences within the channels or cages \cite{Vellamarthodika2022, Jobic_zeolite_dynamics, Leroux_SO2, Dhiman2023, Vietze1998}.
Endohedral fullerenes offer a particularly clean example of quantum confinement: molecules such as H$_2$ and CH$_4$ confined inside C$_{60}$ exhibit discrete quantum rotational states that are exquisitely sensitive to cage symmetry and weak van der Waals interactions \cite{Carravetta_fullerenes, Horsewill_rotational, Xu2013}. These systems provide compelling evidence of how confinement and symmetry shape quantum rotational behavior across a wide range of host architectures.

In gas-filled ice structures under high pressure \cite{Bove2019}, the extreme confinement imposed by the ice hosting lattice leads to the suppression of molecular rotations even at high temperatures, enabling the onset of orientational ordering of guest molecules  \cite{Okuchi2007, Schaack2018, Tse2024}. In certain cases, this ordering is accompanied by molecular distortion \cite{Schaack2018}, and can even trigger symmetry-lowering transitions in the host lattice itself \cite{Schaack2019, Andriambariarijaona2025}. These phenomena underscore the unique interplay between quantum confinement and structural flexibility in filled-ice systems—an interplay rarely observed in classical porous materials.

Among these, hydrogen hydrates (HH)—where hydrogen molecules are encapsulated within an extended water lattice—stand out as a particularly rich system for exploring host–guest interactions under extreme thermodynamic conditions. HH comprises five well-characterized phases. The first, i.e., the clathrate structure sII, contains hydrogen molecules trapped in large polyhedral cages within a hydrogen-bonded water network \cite{Lokshin2004, ranieri_quantum_2019}. The other four—C0, C1, C2, and C3—are filled-ice structures in which hydrogen molecules occupy interstitial voids within compact ice-like frameworks. Specifically, C0 adopts the open-channel ice XVII topology \cite{DelRosso2021}, C1 corresponds to the layered ice II structure \cite{Carvalho2021}, while C2 and C3 are based on the cubic ice (Ic) framework \cite{ranieri_observation_2023}.

While sII is only stable at cryogenic temperatures, both C1 and C2 persist at room temperature under high pressure. C1 is stable between 0.9 and 2.7 GPa (and metastable up to 5.2 GPa), whereas C2 stabilizes above 2.7 GPa and transforms into C3 at temperatures above 670 K and pressures above 30 GPa. C3 remains stable up to at least 90 GPa and can be recovered metastably down to 7 GPa \cite{Andriambariarijaona2025}.

Among these phases, C2 is particularly intriguing. It adopts a topology analogous to that of ice VII, consisting of two interpenetrating but non-interconnected diamond-like sublattices. However, unlike ice VII—where both sublattices are composed of water molecules—in C2, one sublattice is entirely formed by hydrogen molecules. This breaks the structural equivalence, lowers the overall symmetry, and results in a unique dual-framework architecture. Remarkably, C2 exhibits the shortest known H$_2$–H$_2$O distances among hydrogen hydrates, and its 1:1 stoichiometry gives rise to exceptionally strong host–guest interactions.

Under such extreme densities confinement effects predominate, affecting not only the hydrogen molecules within their sublattice but also the protons constituting the ice framework. As pressure increases, the distinction between host and guest progressively diminishes, and the system is more appropriately described as two equivalent, interpenetrating quantum lattices of distinct species: H$_2$ and H$_2$O.

\begin{figure}[]
    \centering
    \includegraphics[width=1\linewidth]{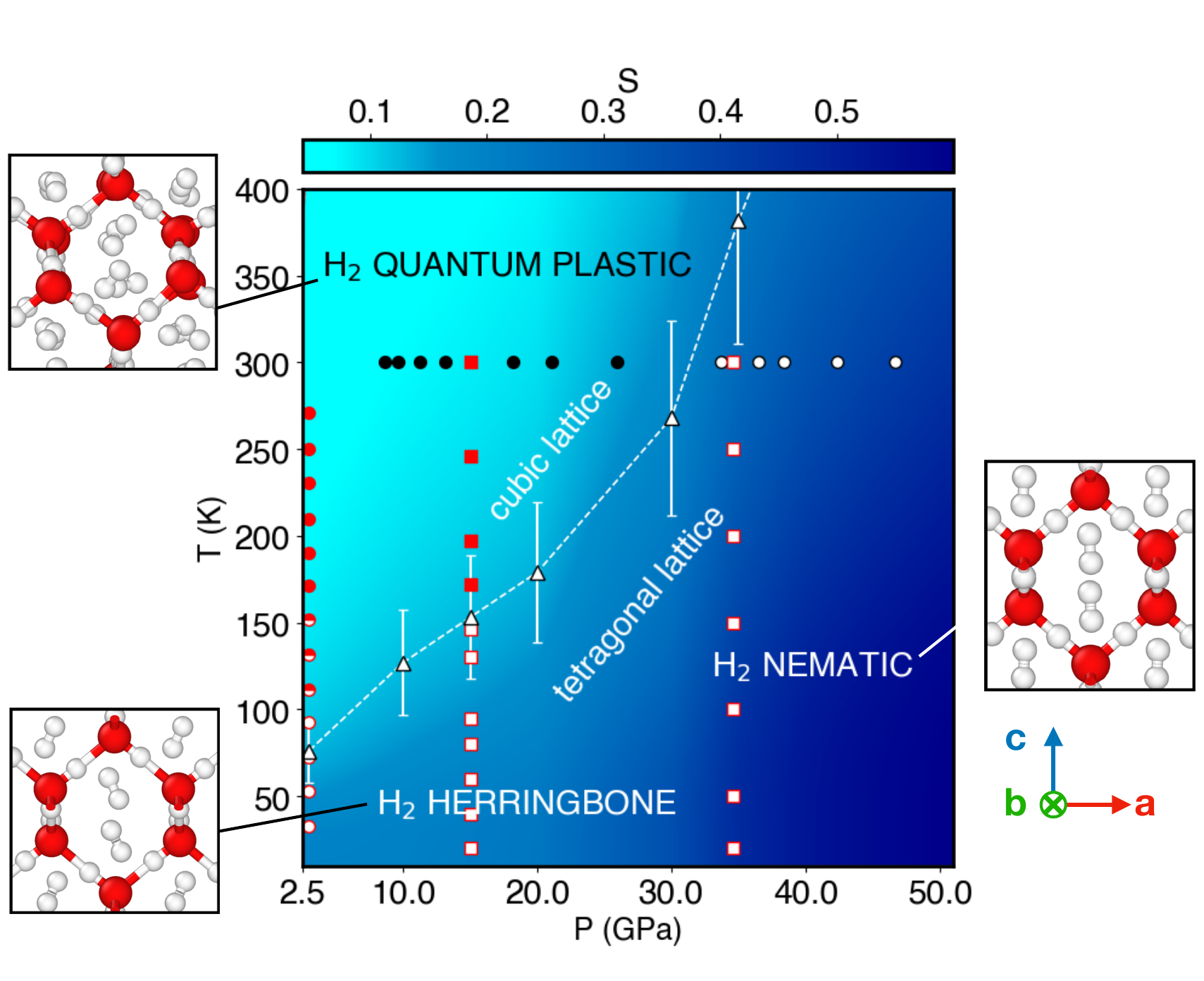}
    \caption{Pressure-Temperature phase diagram of the C2 hydrogen hydrate as derived by our quantum embedded calculations and experimental data. Some experimental pattern are indicated in Figure : XRD data are shown as circles, colored red or black depending on whether they belong to the temperature- or pressure-ramp series; raman measurements are indicated by squares. The summary of experimental thermodynamic path is reported in Table \ref{tab:experimental_conditions}. Symbols are filled in the cubic phase and open in the tetragonal phase, while half-filled symbols indicate the transition between them. The background color represents S, the orientation factor (defined in Methods) of H$_2$ computed in the quantum embedded framework.
    The white triangles with their associated error bars are derived from our molecular dynamics simulations, utilizing the orientation factor as detailed in the Methods section. The dashed lines serve as a guide to the eye to locate the cubic to tetragonal transition. The three insets display the optimized geometries obtained via DFT (0 K) for the quantum plastic phase (upper left), herringbone phase (lower left), and nematic phase (right) of H$_2$ orientational states. }
    \label{fig:PhaseDiagram}
\end{figure}

The rotational dynamics of hydrogen in hydrogen-filled ices are strongly pressure- and temperature-dependent. Increasing pressure and lowering temperature progressively restrict guest rotation, inducing stepwise orientational ordering \cite{Hirai2006}, analogous to effects seen in solid methane \cite{Bini1997, Bykov2021} and solid hydrogen \cite{mao1994ultrahigh, Hemley1996, Silvera2021}, where intermolecular interactions produce roton band splittings. Similar but stronger effects occur in hydrogen clathrates \cite{Strobel2009}.

In the C2 phase, first-principles calculations predicted a pressure-induced cubic-to-tetragonal distortion at low temperature via hydrogen alignment along the c-axis \cite{Zhang2012}, confirmed in D$_2$O–D$_2$ by XRD peak splitting and roton band splitting above 20 GPa at 10 K \cite{Hirai2012}. Additional high-pressure phases (HH-HP1 at ~40 GPa and HH-HP2 at ~60 GPa) involve further orientational ordering and possible hydrogen-bond symmetrization \cite{Machida2008, Hirai2012, Machida2010}.

Our recent work refined this phase diagram: we observed tetragonal distortion of C2 above 30 GPa at room temperature, amorphization above 60 GPa, and transformation to C3 upon laser heating \cite{ranieri_observation_2023, Andriambariarijaona2025}. We also determined that hydrogen-bond symmetrization occurs at ~26 GPa—much lower than in ice VII—due to shorter O–O distances in the cubic ice Ic framework \cite{Monacelli2025}. 

In the present work, we go beyond these macroscopic observations to elucidate their microscopic origin: we demonstrate that hydrogen-bond symmetrization in the water lattice triggers the nematic alignment of the H$_2$ sublattice, and that this collective orientational ordering, in turn, drives the tetragonal distortion of the host framework. By disentangling this sequence of coupled events, we establish the microscopic mechanism linking proton symmetrization, guest orientational order, and host lattice deformation in C2. We probe these phase transitions via combined \textit{ab initio} simulations, low-temperature Raman spectroscopy, and synchrotron XRD \cite{Aoki96, Gonchy1996, pruzan_raman_1997, Komatsu2023, Poreba_2023}, highlighting how enhanced host–guest coupling under pressure suppresses rotational freedom and stabilizes orientational order at much lower pressures than in pure hydrogen.

\section*{Results and Discussion}
Figure \ref{fig:PhaseDiagram} presents the pressure–temperature phase diagram constructed from our combined computational and experimental approaches, with the thermodynamic paths probed experimentally summarized in Table \ref{tab:experimental_conditions}. Within the C2 phase, H$_2$ exhibits three distinct regimes, dictated by temperature and the degree of coupling with the surrounding water lattice.

\begin{table}[]
    \centering
    \begin{tabularx}{\columnwidth}{p{1.5cm}XXp{2cm}}
    \hline
    \rowcolor[HTML]{ECF4FF} 
                                       & \textbf{XRD}           & \textbf{Raman}          \\ \hline
\multicolumn{1}{l|}{\textbf{Isobar}}   & 2.45(3) GPa               & 3.5 GPa                 \\
\multicolumn{1}{l|}{}                  & 300 $\rightarrow$ 30 K & 300 $\rightarrow$ 100 K \\
\multicolumn{1}{l|}{}                  &                 & 14.7 GPa                  \\
\multicolumn{1}{l|}{}                  &                        & 300 $\rightarrow$ 5 K   \\
\multicolumn{1}{l|}{}                  &                        & 35 GPa                  \\
\multicolumn{1}{l|}{}                  &                        & 300 $\rightarrow$ 20 K  \\ \hline
\multicolumn{1}{l|}{\textbf{Isotherm}} & 300 K                  & 300 K                   \\
\multicolumn{1}{l|}{}                  & 2 $\rightarrow$ 46 GPa & 10 $\rightarrow$ 40 GPa \\ \hline
    \end{tabularx}
    \vspace{5pt}
    \caption{Summary of experimental thermodynamic paths for X-Ray Diffraction (XRD) and Raman.}
    \label{tab:experimental_conditions}
\end{table}

At high temperatures and pressures below 30 GPa (above the dashed line) the H$_2$ sublattice in H$_2$ hydrate realizes a quantum plastic crystal. In this state, the H$_2$ molecules are ordered on a diamond-type lattice, defining well-localized translational equilibrium sites, yet their orientations remain dynamically disordered \cite{Rescigno2025}. Each H$_2$ behaves as a nearly free quantum rotor, i.e., the rotational degrees of freedom are not frozen but remain active down to relatively low temperatures due to quantum fluctuations. This combination of translational order and orientational quantum mobility is the defining feature of a quantum plastic phase. In contrast, the water sublattice is proton-disordered but not plastic, since the orientations of H$_2$O molecules are constrained by the ice rules and do not undergo free rotational dynamics. The overall symmetry of the crystal is cubic, consistent with previous DFT predictions and experimental observations \cite{ranieri_observation_2023, Zhang2012, Andriambariarijaona2025}.

Upon crossing the dashed boundary, either through temperature reduction or compression, H$_2$ molecules undergo orientational ordering. At high pressure this results in a phase where H$_2$ are  aligned along the c axis of the crystal. This ordering transition can be quantified by the orientation factor $S$ (See Methods) computed based on our quantum embedded calculations as described in the following, and visualized as a color gradient in Fig. \ref{fig:PhaseDiagram}. The factor $S$, which ranges from 0 (complete orientational disorder) to 1 (perfect alignment), measures the collective alignment of the hydrogen molecules along the 
$c$ axis of the crystal, capturing both the orientational order imposed by the lattice geometry and the intrinsic thermal quantum fluctuations encoded in the density matrix.

Within the orientationally ordered phase—occurring below 30 GPa and 100 K—H$_2$ molecules adopt a herringbone-like arrangement, as shown in the inset of Fig. \ref{fig:PhaseDiagram}. This configuration favors alignment along the crystalline $c$ axis. The tilt angle of the molecules relative to the $c$ axis varies continuously with pressure, reflecting the energetically favorable response to lattice compression at low temperatures. As pressure increases, the tilt angle gradually decreases, approaching full alignment along the $c$ axis, i.e the nematic phase. This progressive reorientation is also captured by the pressure dependence of the orientation factor $S$ in Fig. \ref{fig:PhaseDiagram}.

The low-temperature, high-pressure phases ($<$ 50 GPa) of H$_2$ and D$_2$ are unique molecular systems in which free rotations are governed by both quantum effects and temperature \cite{VanKranendonk1968}. Molecular hydrogen occurs in two nuclear-spin isomers, both partially occupied at ambient conditions. When the proton spins couple to the singlet state with total nuclear-spin quantum number $\mathcal{J} = 0$, defined by $\mathbf{\mathcal{J}} = \mathbf{\mathcal{J}}_1 + \mathbf{\mathcal{J}}_2$, where $\mathcal{J}_1 = \mathcal{J}_2 = \tfrac12$ are the proton resolved nuclear spins, the resulting para-hydrogen possesses an antisymmetric spin wavefunction and, to satisfy overall fermionic antisymmetry, its spatial part is restricted to even rotational quantum numbers $l$. Coupling the same proton spins to the triplet state ($\mathcal{J} = 1$) produces ortho-hydrogen, whose symmetric spin wavefunction permits only odd $l$ values.

At low pressures, Raman spectra of solid hydrogen and clathrates \cite{Cooke2020,Strobel2009,Strobel2011,ranieri_observation_2023,ranieri_large-cage_2024} display free-rotor peaks with energies $E_l = \frac{\hbar^2}{2I}l(l+1)$ 
where $I$ is the moment of inertia of the rotor and $l$ the angular momentum, with selection rule $\Delta l = \pm 2$. In the free molecule, rotational levels are $(2l+1)$-fold degenerate, and the main gas-phase Raman bands are $S_0(0)$  ($l = 0 \rightarrow 2$) (354 cm$^{-1}$) and $S_0(1)$ ($l = 1 \rightarrow 3$) (587 cm$^{-1}$) \cite{Cooke2020,Pena-Alvarez2020}. In the clathrate phase and in low-pressure filled ice C0 and C1 phases, the rotor bands remain close to the free-rotor values and their degeneracy is preserved. In the high-pressure C2 phase, however, the anisotropic crystal field lifts this degeneracy and shifts the main ortho–para transition band by 4 meV—about 30\% of its value \cite{dicataldo2024}. With increasing pressure, these bands broaden, making individual rotational levels increasingly difficult to resolve.

\begin{figure}[h!]
\centering
\includegraphics[width=\textwidth]{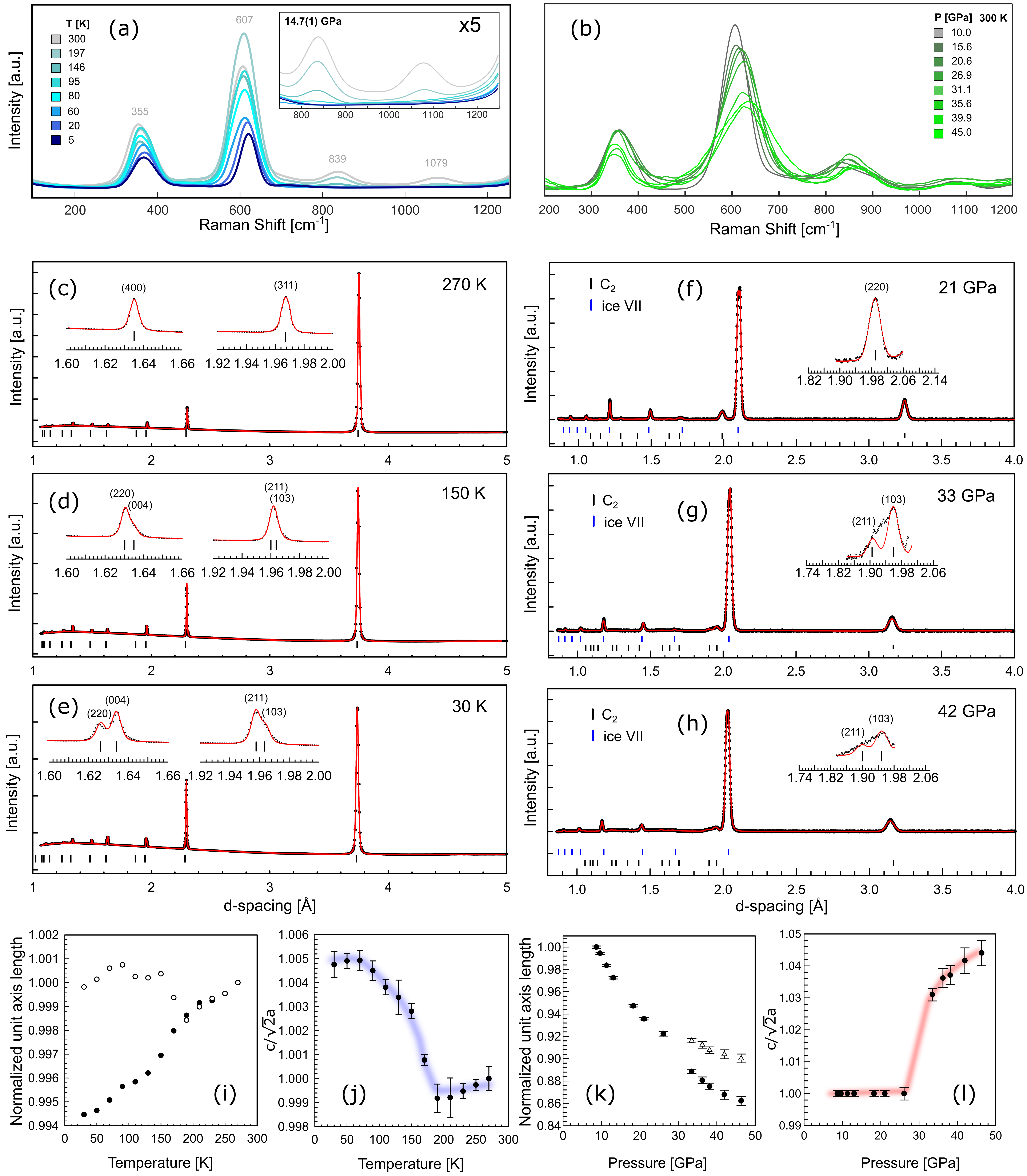}
\caption{Raman spectra as a function of temperature at P= 14.7(1) GPa (a) and as a function of pressure at T=300 K (b). Represnetative X-ray diffraction data as a function of temperature at P=2.45(3) GPa (left panel,c-e)  and as a function of pressure at T=300 K (right panel, f-h). The extracted evolution of the lattice parameters(in the tetragonal setting) and the degree of asymmetrization upon cooling (i,j) and compression (k,l),  are plotted in the bottom panels, respectively.} 
\label{fig:XRD and Raman}
\end{figure}

Figure \ref{fig:XRD and Raman}a,b shows the evolution of the Raman intensity distribution among the $S_x(0)$ transitions in a gas-loaded hydrogen hydrate sample (see Methods) as a function of temperature and pressure. XRD confirmed phase-pure C2 after gas loading, whereas cryo-loaded samples contained 83 mol\% H$_2$O due to decomposition of the sII substrate (6:1 H$_2$O:H$_2$ stoichiometry).

At low temperatures, the $S_{0}(2)$ and $S_{0}(3)$ bands weaken and vanish below 150 K due to thermal depopulation of higher rotational levels; in pure hydrogen, these disappear at 120 K (3 GPa) \cite{Gregoryanz2018}. In the compressed hydrate, the Raman evolution is more complex: a sudden broadening and asymmetry around 30 GPa suggest a change in the rotational state of encapsulated H$_2$ \cite{Eggert1999, Cooke2020}, likely driven by modifications to the crystalline field from the surrounding water framework during its transition to a less compressible phase \cite{Andriambariarijaona2025}. The instability of C2 above 20 GPa toward decomposition into ice VII and molecular hydrogen \cite{Machida2010} may also contribute to inhomogeneous Raman signals from free H$_2$. The intensity oscillations reported near 20 GPa \cite{Machida2010} were not observed here; spectra were consistently collected from the same spot after each pressure step, with a 30-minute equilibration, ensuring that the observed increases in the $S_0(1)$/$S_0(0)$ intensity ratio upon cooling (14.7 GPa) and compression (300 K) reflect enhanced ortho–para conversion. Both high pressure and low temperature are known to facilitate this process in solid hydrogen \cite{Eggert1999, Cooke2020}.

XRD measurements reveal that either cooling below 150 K at 2.45 GPa or compressing above 27 GPa at RT induces peak splitting (Fig. \ref{fig:XRD and Raman}c–h), signaling a cubic ($Fd\bar{3}m$) to tetragonal ($I4_1/amd$) transition. Although the splitting pattern is similar in both cases, it is markedly sharper and an order of magnitude larger under compression. The pressure-driven transition begins abruptly at around 30 GPa, coinciding with the onset of Raman anomalies—evidence of coupling between H$_2$ rotational dynamics and framework distortion. In both scenarios (cooling and pressurization), the $c$ axis stiffens while the orthogonal directions contract (Fig. \ref{fig:XRD and Raman}i,k). The distortion, quantified as $c/\sqrt{2}a$, increases smoothly, reaching 0.5\% at 30 K (2.45 GPa) and 4.6\% at 48 GPa (300 K) (Fig. \ref{fig:XRD and Raman}j,l).

\begin{figure}[h!]
    \centering
    \includegraphics[width=1\textwidth]{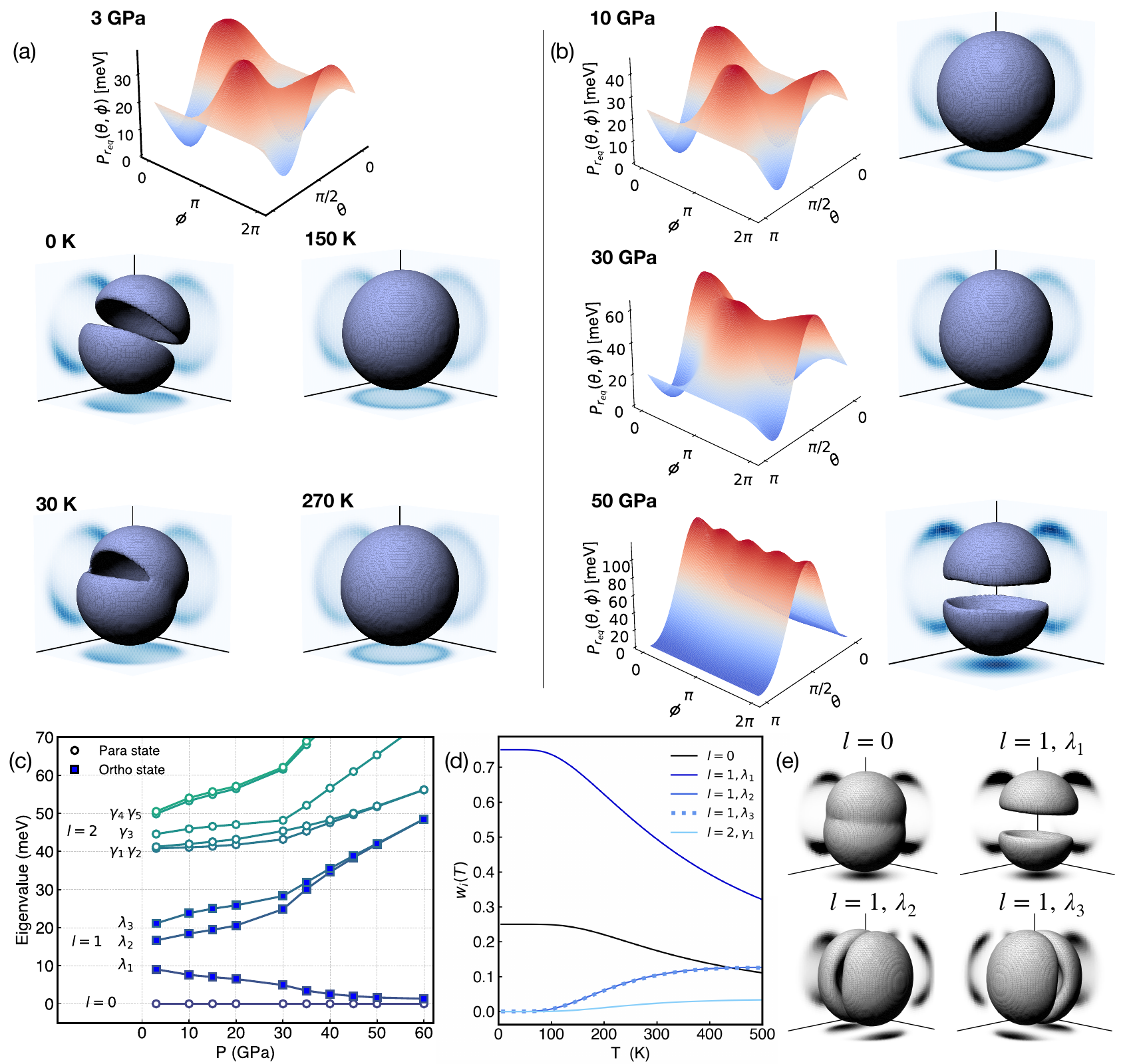}
    \caption{Isosurfaces of the density $\rho(\vec{r})$ and angular potential $P_{r_{eq}}(\theta, \phi)$ evaluated at the equilibrium H$_2$ bond length, shown as a function of temperature at $P = 3$ GPa (a), and as a function of pressure at $T = 300$ K (b). The density is projected onto the three planes ($xy$, $yz$, and $xz$) and normalized with respect to the maximum density at 0 K, 3 GPa for panel (a), and at 300 K, 50 GPa for panel (b).
    Pressure dependence of the first nine eigenvalues at $T = 0$ K. Ortho and para levels are shown as circles and squares, respectively (c).
    Boltzmann weights vs. temperature at P = 50 GPa for the first five eigenvalues (d).
    Isosurfaces of the wavefunctions squared modulus for the ground state (para, $l = 0$) and the first three excited states (ortho, $l = 1$) of the H$_2$ molecule at P = 50 GPa (e).}
    \label{fig:simulation sum}
\end{figure}

To probe the molecular-scale mechanism linking proton alignment to the lattice distortions seen in XRD, we performed simulations of a crystal-embedded H$_2$ molecule by solving the Schrödinger equation with  an effective potential corresponding to the crystal field generated by the water framework and surrounding H$_2$ molecules. Figure \ref{fig:simulation sum}(a–b) shows the evolution of H$_2$ density isosurfaces with temperature and pressure, along with the angular component $P_{r_{eq}}(\theta, \phi)$ of the external potential $V_{ext}(r, \theta, \phi) = R(r) + P_r(\theta, \phi)$ at the equilibrium H$_2$ bond length $r=r_{eq}$. $\theta$ and $\phi$ are respectively the polar and azimuthal angles of the molecular orientation in a cubic lattice reference. As described in Methods and Supporting Information, densities at a given $T$ are obtained from the para and ortho thermal density matrices mixed together with a fixed 1:3 para–ortho ratio.

At low pressure ($P=3$ GPa, panel \ref{fig:simulation sum}(a)), the external angular potential exhibits minima that force the molecules into a herringbone-like arrangement. As $T$ increases within the C2 stability range, para- and ortho-wavefunctions mix and the density approaches spherical symmetry. With increasing pressure, intermolecular interactions intensify and potential barriers rise, while the minima shift toward $\theta = 0 \ [\pi]$, ultimately favoring full nematicity. At room temperature (panel \ref{fig:simulation sum}(b)), the density evolves from nearly spherical shells at 10 GPa—characteristic of the quantum-plastic phase—to well-defined lobes aligned with the $c$-axis at 50 GPa.

The eigenvalue evolution shown in Fig.\ref{fig:simulation sum}(c) demonstrates how the anisotropic crystal field in the C2 phase lifts the free-rotor degeneracy and induces a pressure-dependent giant splitting, in agreement with earlier findings \cite{dicataldo2024}. This quantum mechanical picture explains the emergence of nematic ordering at high pressures and its persistence to elevated temperatures. At low pressure, within the C2 stability range, the small energy gap between the fundamental para state ($l=0$) and the three ortho states ($l=1, \lambda_1, \lambda_2, \lambda_3$) enables thermal mixing above $\sim$100 K, yielding nearly spherical densities. Increasing pressure drives a sharp rise in the splitting between $\lambda_1$ and $\lambda_{2,3}$, bringing the $\lambda_1$ level close to $l=0$. This restricts significant mixing to $l=0$ and $l=1, \lambda_1$, both oriented along the $c$ axis, producing a nematic density distribution. The fixed 1:3 para–ortho ratio further enhances the statistical weight of $\lambda_1$, stabilizing orientational order up to $\sim$500 K (Fig.\ref{fig:simulation sum}d–e).


\begin{figure}
    \centering
    \includegraphics[width=1\textwidth]{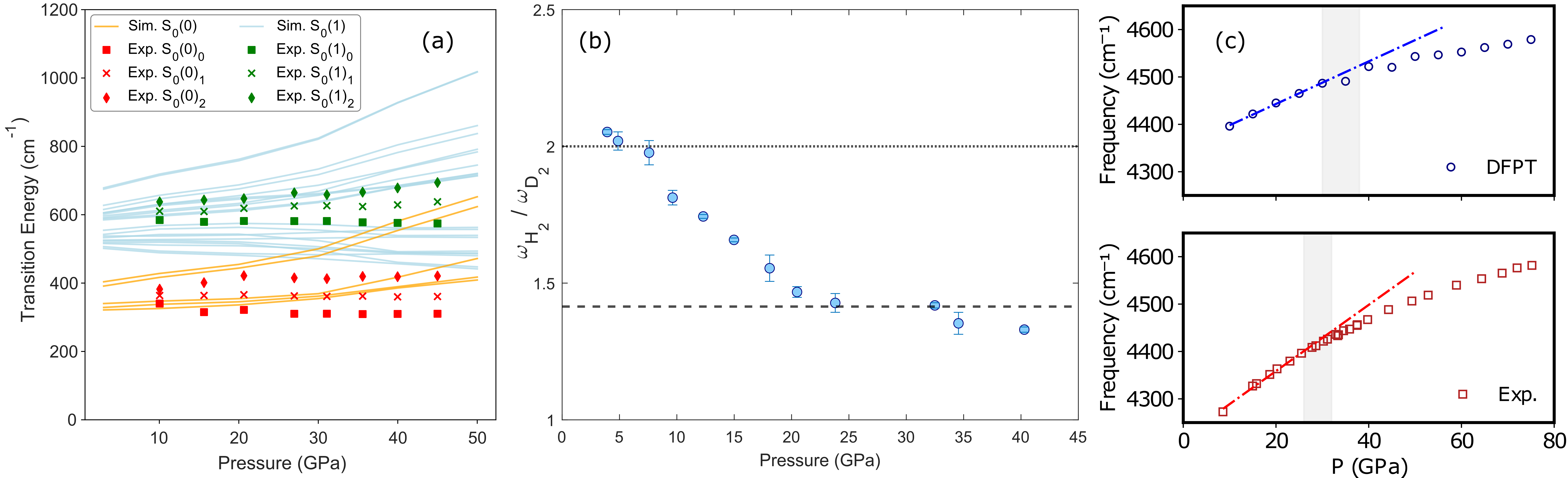}
    \label{fig:figure5a}
    \hfill
    \caption{(a) Comparison of the experimental Raman points and all simulated S$_0$(0) and S$_0$(1) rotons. The blue and orange lines are theoretical whilst the red and green point indicate the S$_0$(0) and S$_0$(1) rotons respectively. The rotons were modelled with three components each, as indicated by the square, crosses and diamonds (b) Pressure dependence of the S$_0$(1) frequency ratio $\omega_{H_2}/\omega_{D_2}$ derived from the experimental Raman data. The rigid rotor ratio is indicated by the dotted line at $\omega_{H_2}/\omega_{D_2} = 2$, whilst the quantum harmonic oscillator ratio is indicated by the dashed line at $\omega_{H_2}/\omega_{D_2} = \sqrt{2}$. HH crosses into the harmonic oscillator regime at around 27 \,GPa, the same pressure a structural change is previously reported \cite{Andriambariarijaona2025} (c) Experimental Raman frequencies derived from the fitting procedure described in Methods (squares) and computed Raman frequencies obtained via DFPT (circles) for the hydrogen vibron upon compression of the hydrogen hydrate.
    The change in slope, highlighted by the shadow area, corresponds to the pressure where the hydrogen molecules order.} 
    \label{fig:HH vibron}
\end{figure}

\begin{figure}[t!]
    \centering
    \includegraphics[width=1\textwidth]{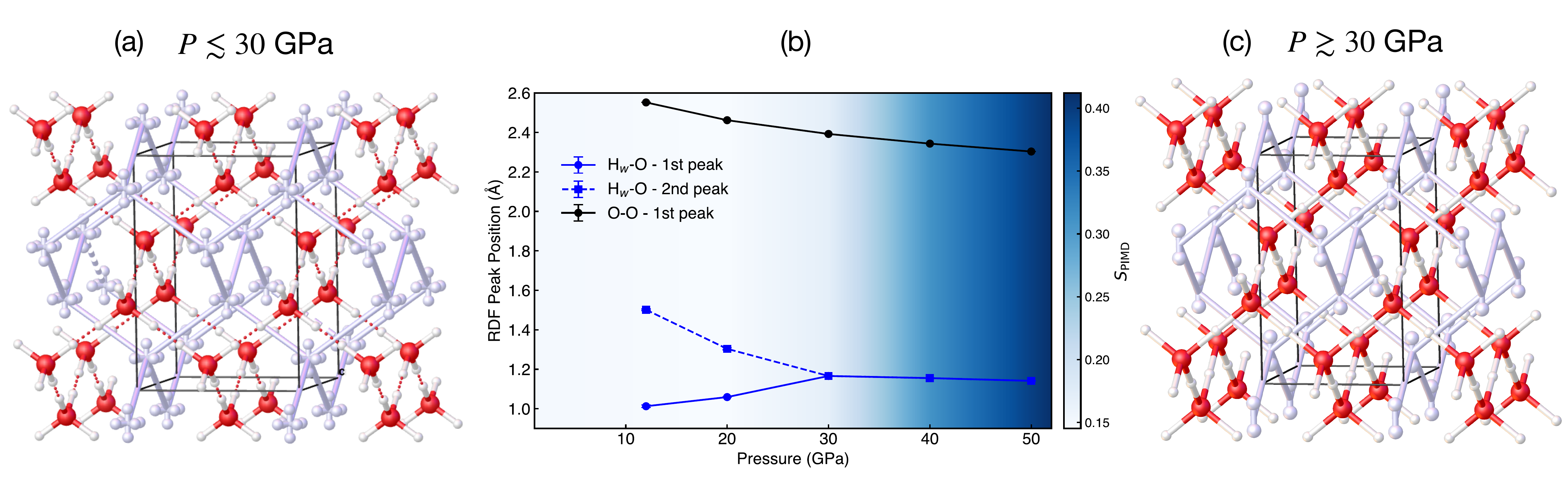}
    \caption{(a) Crystal structure of C2 hydrogen hydrate at low pressures ($P \lesssim 30$ GPa), showing the two interpenetrating cubic sublattices of $H_2O$ and $H_2$. The water–proton network is disordered, while the $H_2$ molecules are orientationally disordered at high temperature and adopt a herringbone-like arrangement at low temperature.
    (b) Pressure dependence of the first (solid lines) and second (dotted lines) peak positions in the radial distribution functions $g(r)$ for the $H_w–O$ and $O–O$ pairs, from PIMD simulations at $T = 300$ K. The background color gradient represents the orientation factor $S_{\text{PIMD}}$ extracted from those simulations.
    (c) Crystal structure of C2 hydrogen hydrate in the symmetrized (water–proton) state at high pressures ($P \gtrsim 30$ GPa). At the highest pressures, $H_2$ molecules become nematic along the $c$-axis, coinciding with the tetragonal distortion direction.
    }
    \label{fig:final}
\end{figure}

From these computed eigenvalues, we can determine the Stokes rotational transitions $S_0(0)$ and $S_0(1)$, plotted in Fig.~\ref{fig:HH vibron}(a) as orange and blue curves, respectively. The comparison with the experimental Raman data in the same panel shows good agreement, particularly in the pressure evolution of the main features, thus validating the simulation framework. The pronounced change in slope of both $S_0(0)$ and $S_0(1)$ near 30 GPa coincides with the cubic–tetragonal distortion observed in XRD and with the calculated onset of dominant $\lambda_1$ – $l=0$ mixing (Fig.\ref{fig:simulation sum}c-d), supporting the interpretation that the structural transition is driven by a pressure-induced orientational ordering of the guest molecules.

Panel \ref{fig:HH vibron}(b) further supports this picture by showing the pressure dependence of the ratio between the $S_0(1)$ Raman frequencies measured for an H$_2$-hydrate ($\omega_{\mathrm{H}_2}$) and for a D$_2$-hydrate ($\omega_{\mathrm{D}_2}$). At low pressure, the ratio remains close to the free-rotor limit of 2, reflecting the $1/I$ scaling of rotational energies with the molecular moment of inertia. Upon compression, the increasingly strong crystalline field exerted by the rigid water lattice progressively hinders the molecular rotations, effectively locking the H$_2$/D$_2$ orientation relative to the host cages. As rotational freedom is quenched, the excitations acquire librational character and the ratio approaches the harmonic mass-scaling limit $\sqrt{2}$, characteristic of an oscillator with frequency $\propto m^{-1/2}$. A qualitatively similar evolution has been reported for pure solid hydrogen and deuterium at much higher pressures (above $\sim$140 GPa) \cite{Pena-Alvarez2020, Cooke2020}, and a direct comparison between the dispersion curves for the hydrate and for the elemental solids is provided in the Supplementary Information (Fig.~\ref{fig:RamanH2D2}). In analogy with the dense elemental phases, if the hindering potential in the hydrate is sufficiently anharmonic — i.e. weaker than purely harmonic at large angular displacements — the ratio can transiently overshoot below $\sqrt{2}$ before converging, reflecting an enhanced softening of the librational mode. Notably, the roton–libron crossover in the hydrate occurs at pressures an order of magnitude lower than in pure hydrogen/deuterium, highlighting the much stronger orientational constraints imposed by the host lattice. The observed trend thus constitutes direct spectroscopic evidence for the progressive locking of the guest-rotor orientation due to the strengthening of H$_2$/D$_2$–H$_2$O coupling under pressure.

This progressive hindering of the guest rotations revealed in panel \ref{fig:HH vibron}(b) has a direct counterpart in the behavior of the intramolecular H–H stretching mode, since changes in rotational freedom modify the coupling between rotational and vibrational degrees of freedom and thus the vibron pressure dependence. Consequently, from the analysis of the H–H stretching mode in Fig.~\ref{fig:HH vibron}(c) comparing Density Functional Perturbation Theory (DFPT) calculations with experimental Raman frequencies it turns out that both simulation and experiment display a clear change in slope near 30 GPa. This is consistent with the signatures seen in the rotational modes and further linking the vibron behavior to the onset of pressure-induced orientational ordering predicted by our quantum embedded model.

The structure of the C2 hydrogen hydrate at low pressure is shown in Fig.\ref{fig:final}(a). The two interpenetrating hydrogen and water cubic sublattices are fully orientationally disordered: the protons in the water framework are randomly distributed between equivalent hydrogen-bond configurations, and the H$_2$ molecules adopt random orientations within their cages. Panel \ref{fig:final}(b) reports the evolution with pressure of the positions of the two oxygen atoms closest to a given water proton H$_w$—the intramolecular oxygen covalently bound to H$_w$ and the oxygen of the neighboring molecule to which H$_w$ is hydrogen-bonded—shown in blue, as extracted from our Path-Integral Molecular Dynamics (PIMD) trajectories (see Methods). With increasing pressure, these two nearest oxygen atoms approach to each other and their corresponding distances to H$_w$ converge at $\sim$26 GPa \cite{Monacelli2025}, indicating proton symmetrization and the formation of symmetric $O-H\cdots O$ bonds. The first peak position of the oxygen–oxygen ($O$–$O$) radial distribution function is shown in black, while the background color encodes the orientation factor $S_{\text{PIMD}}$ of the H$_2$ molecules, computed from path-integral molecular dynamics as an ensemble average over beads and time steps, as defined in Methods. This quantity implicitly incorporates both the geometrical arrangement of the molecules and their thermal-quantum fluctuations relative to the $c$ axis. Before symmetrization, compression primarily shortens the hydrogen bonds, with little ordering of the H$_2$ sublattice. Proton symmetrization abruptly stiffens the $O-H\cdots O$ network, producing a sharp increase in the bulk modulus $B$ \cite{ranieri_observation_2023, Andriambariarijaona2025} and making the ice sublattice much less compressible. Beyond this point, further isotropic compression is energetically disfavored; instead, it becomes more favorable to reorient the H$_2$ molecules with their bond axes parallel to the $c$ direction [panel \ref{fig:final}(c)], reducing the volume along $a$ and $b$ owing to the increased compressibility along these axes. This collective alignment of the H$_2$ sublattice drives the tetragonal elongation of the H$_2$O framework and produces a second, more gradual rise in $B$ \cite{ranieri_observation_2023, Andriambariarijaona2025}. In other words, the lattice first complies with increased pressure through hydrogen-bond rigidification in the water network, and then through anisotropic reorientation of the hydrogen network, rather than by further reducing the $O-O$ separation. Figure~\ref{fig:final} thus reveals a two-stage process in which the quantum-mechanical behavior of the water and hydrogen sublattices becomes tightly interlocked, their structural and dynamical degrees of freedom evolving in unison and merging into a single quantum lattice.

Our results demonstrate that C2 hydrogen hydrate undergoes a pressure-driven, two-stage transformation in which the hindering of the rotational dynamics of the hydrogen sublattice becomes intertwined with the distortion of the ice sublattice. The first stage—proton symmetrization in the host lattice—abruptly stiffens the hydrogen-bond network, while the second—anisotropic alignment of the H$_2$ molecules—relieves further compression and induces a tetragonal distortion. This collective alignment is directly manifested in the structural transition of the lattice, marking the point where the guest reorientation and host deformation proceed in lockstep. The coupling is evident across structural, elastic, rotational, and vibrational observables, linking the roton–libron crossover of the guest molecules to the proton ordering of the host. In this regime, the motions of the host lattice and the guest molecules become strongly interconnected, forming collective quantum modes rather than distinct subsystems. This provides an exquisite  example of coupled nuclear quantum dynamics in a crystalline solid, with potential implications for other hydrogen-rich materials under high pressure.

\section*{Methods}
\subsection*{Experimental Details}
Samples were prepared in two different ways, depending on the specific experiment, cryo-loaded or gas-loaded in the diamond anvil cell (DAC). In the first case, polycrystalline hydrogen hydrate samples with the sII clathrate structure were prepared by exposing H\textsubscript{2}O ice (I$h$) to gas H\textsubscript{2} at 0.28 GPa for 20–30 min, following the method described in ref. \cite{ranieri_quantum_2019}. The ice was made of spheres with typical diameters of a few tens of micrometers. After preparation, the samples were recovered at ambient pressure and stored at liquid nitrogen temperature. Then, a small amount of sample was loaded at liquid nitrogen temperature into a diamond anvil cell partially immersed in a liquid nitrogen bath and in an atmosphere saturated with nitrogen vapor to avoid water or oxygen condensation on the culet. The loaded samples were then compressed to pressures of a few GPa before being warmed to room temperature.
This is a similar loading procedure to what we had previously done for methane hydrate \cite{Schaack2018,Schaack2019}.
This procedure inevitably produces a mixture of C2 hydrogen hydrate and excess pure ice, due to the difference in molar ratios between the clathrate sII phase and the C2 phase.
Clathrate sII samples prepared following our method have a molar ratio H\textsubscript{2}O:H\textsubscript{2} between 4 and 5, as we have verified by neutron diffraction in the past \cite{ranieri_quantum_2019}, and the C2 phase is characterized by a molar ratio of 1.
Two samples were prepared directly in the DAC by loading room-temperature liquid H\textsubscript{2}O and high-pressure H\textsubscript{2} using a gas loading setup. Briefly, liquid water was loaded together with an air bubble, consisting of approximately 70\% of the sample chamber volume. The air bubble was replaced with 99.99+\% H\textsubscript{2} at 1400 bar, and the DAC was subsequently pneumatically closed and pressurized. These samples were measured by both Raman spectroscopy and by XRD.
Some of the cryoloaded samples contained a small amount of pure N\textsubscript{2} and/or pure O\textsubscript{2}, both of which can easily be identified from their respective vibron Raman peaks. They were trapped in the sample chamber during cryoloading and would have acted as a pressure-transmitting medium. In the gas-loaded samples, H\textsubscript{2} itself was acting as the pressure transmitting medium.
Culet diameters ranging between 200–500 µm for the IIas quality diamond anvils were used. Re-foil gaskets were used to contain the sample whilst ruby and/or gold spheres were used as pressure calibrants in Raman and X-ray experiments respectively.

Angle-dispersive X-ray powder diffraction patterns were acquired at ID15b at ESRF (Grenoble, France) using a monochromatic X-ray beam ($\lambda$=0.41 \AA) and an Eiger 2 9M CdTe flat-panel detector, with a typical acquisition time of a ten seconds.\cite{Poreba_2023}
The beam spot size was 6×6 µm, and the DAC was rocked 5° during the acquisition.
The 2D diffraction patterns were treated to mask the Bragg peaks of the diamond anvils, then integrated into one-dimensional patterns, and a smooth polynomial background was subtracted. Le Bail refinements were performed using Topas Academic version 6 \cite{Coelho2018}.
In the XRD measurements, the pressure was determined using the equation of state of gold from ref. \cite{Fei2007}.
Raman spectra at ambient temperature were acquired using a commercial Horiba Jobin-Yvon LabRam HR800 Raman spectrometer in a backscattering geometry, equipped with a COBOLT SambaTM green 532 nm laser, with a nominal power of 1000 mW. Low-temperature Raman spectra were collected  using an Argon laser (120 mW output power) tuned to 514.32 nm and a Peltier-cooled  HR460 spectrograph with 1500 gr/mm grating.
The acquisition time was typically about a minute.
Reference spectra were recorded at each pressure from the gasket close to the sample chamber, and particular care was taken to check the effect of the background subtraction on the fit results.
In the Raman measurements, pressure was determined by the shift of the R1 ruby fluorescence line \cite{Shen2020}, from the edge of the stressed diamond signal (above 40 GPa) \cite{Akahama_2006}, or from the vibron frequency of pure solid H\textsubscript{2} \cite{Howie_2012}.
A  helium cryostat was used for low-temperature measurements.

\subsection*{Computational Details}
\emph{Ab initio} molecular dynamics (AIMD) simulations were carried out on a 2 × 2 × 2 supercell of the C2 phase (space group Pna2$_1$) over a grid of pressures from 3 to 50 GPa and temperatures from 25 to 300 K. Calculations employed the Perdew–Burke–Ernzerhof (PBE) exchange–correlation functional \cite{PhysRevLett.77.3865} and were performed with both VASP \cite{PhysRevB.54.11169} and Quantum Espresso \cite{Giannozzi_2009, 10.1063/5.0005082} packages. The Path-Integral molecular dynamics approach (PIMD) \cite{feynman1979path, 10.1063/1.441588, 10.1063/1.442141, RevModPhys.67.279, Markland2018} was used to efficiently sample nuclear-quantum effects under the desired thermodynamic conditions. Simulations were conducted for temperatures above 25 K, by using the Path-Integral Ornstein–Uhlenbeck Dynamics (PIOUD) algorithm implemented in Quantum Espresso \cite{doi:10.1021/acs.jctc.7b00017}. After performing convergence tests on both the virial and primitive estimators of the quantum kinetic energy, the number of beads ($N_b$) was set to 40 at 300 K and increased to 160 at 25 K for PIMD runs.
For both AIMD and PIMD, a plane-wave cutoff of 100 Ry and a 2 × 2 × 2 k-point mesh were sufficient to achieve convergence in the DFT solution of the electronic problem. Trajectories were obtained with a time-step of 0.5 fs and a total duration of approximately 10 ps for each (P, T) condition. 

Spin isomers such as ortho- and para- hydrogen are not taken into account in molecular dynamics (MD) simulations. Indeed, to access the true low-temperature physics of the C2 phase while retaining quantum effects and accounting for spin isomers, we solved the Schrödinger equation at 0 K for an H$_2$ molecule in an effective potential modeling the induced interactions between H$_2$ and the water lattice. As detailed in the Supplementary Information section \ref{sec:schro}, going beyond previous studies \cite{doi:10.1021/ja103062g, 10.1063/1.2945895, 10.1063/1.2967858, PhysRevLett.133.236101}, we evaluated the full, non-separable potential $V_{\text{ext}}(r, \theta, \phi)$ by performing self-consistent DFT calculations (Quantum Espresso, 300 eV cutoff, 5×4×5 k-mesh) for each configuration generated on a dense spherical grid of bond lengths and orientations \cite{PhysRevLett.133.236101}. These 0 K DFT energies directly capture the interaction of H$_2$ with the water lattice. We used several exchange–correlation functionals and their comparison is shown in Fig.~\ref{fig:quantumlevels}. For consistency with the MD calculations, the main results are obtained using PBE.

The Schrödinger equation has been solved using a basis set of the form $\psi_n^{lm}(r, \theta, \phi) = \chi_n(r) Y_l^m(\theta, \phi)$, where $\chi_n(r)$ is the radial basis of the Morse potential and $Y_l^m(\theta, \phi)$ are the spherical harmonics.
This approach can be viewed as a mean-field model, in which a single H$_2$ molecule is embedded in the field generated by the surrounding water lattice.

The spatial density at finite temperature $T$ was obtained by treating the two nuclear-spin manifolds of $\mathrm{H}_2$ independently.  Assuming that ortho $(\mathcal{J}\!=\!1)$ and para $(\mathcal{J}\!=\!0)$  populations do not interconvert on the time scale of the simulation, the total density reads

\[
\begin{aligned}
\rho(\mathbf r)\; &=
\sum_{i\in\text{ortho}}
     \frac{f_{\mathrm o}\,e^{-\beta\,(E_i-E_0^{\mathrm o})}}{Z_{\mathrm o}}
     \,|\psi_i(\mathbf r)|^{2}
&\
+\;\sum_{j\in\text{para}}
     \frac{f_{\mathrm p}\,e^{-\beta\,(E_j-E_0^{\mathrm p})}}{Z_{\mathrm p}}
     \,|\psi_j(\mathbf r)|^{2}
\end{aligned}
\]

where

$
\begin{aligned}
Z_{\mathrm o}= \sum_{i\in\text{ortho}} e^{-\beta\,(E_i-E_0^{\mathrm o})}, \quad
Z_{\mathrm p}= \sum_{j\in\text{para }} e^{-\beta\,(E_j-E_0^{\mathrm p})},\\
\end{aligned}
$

Here $\beta=(k_{\mathrm B}T)^{-1}$, $|\psi_{n}|^2$ and $E_{n}$ are the eigenfunctions and eigenvalues obtained from the Schrödinger equation evaluated in the relaxed hydrate lattice; $E_0^{\mathrm o}$ and $E_0^{\mathrm p}$ are the ortho and para ground-state energies, and $f_{\mathrm p}:f_{\mathrm o}=1:3$ is the fixed statistical spin ratio. This approach neglects cage deformation effects, as the potential is evaluated using the relaxed lattice geometry.
A key advantage of the embedded quantum model is that it naturally incorporates 
the correct ortho/para mixture of H$_2$. In contrast, PIMD simulations describe 
only the para state (the ground rotational state), while in reality the ortho 
species is more populated than the para one in a 3:1 ratio. Because the ortho 
states are more localized than the para state, this difference has a major impact 
on orientational properties, particularly at the lowest temperatures. Accounting for the ortho/para mixture is therefore 
essential for a quantitatively accurate description of the orientational regimes 
and phase transitions. Another clear advantage of the quantum embedded model is the possibility to explore low-temperature regimes at constant computational effort, compared to PIMD whose effort increases dramatically as temperature is lowered. We therefore adopted the quantum embedded model as viable approach to study the quantum phase diagram of hydrogen filled ice as a function of pressure and temperature.

To characterize the three distinct orientational regimes for H$_2$ (quantum plastic, herringbone-like configurations, and nematicity \cite{ranieri_observation_2023, Zhang2012}) from the quantum embedded model, we 
introduce
the orientation factor
$S$ 
as follows \cite{annurev:/content/journals/10.1146/annurev.bi.47.070178.004131, doi:10.1021/ma00148a028}. 

In analogy with diffraction theory, where the scattering amplitude can be factorized into a lattice contribution and a local form factor, we define the orientation factor as the product of a 
collective geometric term and an intrinsic quantum factor:
\begin{equation}
S(P,T) = S_g(P)\, S_q(P,T).
\label{S}
\end{equation}

The geometric contribution $S_g$ is evaluated on the 0~K equilibrium geometry obtained from DFT calculations at a given pressure $P$. It is defined from the orientation of each molecule $i$, through the angle $\theta_i$ between the crystallographic $c$ axis (preferential axis of alignment) and the molecular axis. Its definition coincides with the nematic order parameter \cite{10.1063/1.471343, de1993physics}:
\begin{equation}
S_g = \frac{1}{N}\sum_{i=1}^{N}\left(\frac{3}{2}\cos^{2}\theta_i-\frac{1}{2}\right).
\label{Sg}
\end{equation}
This factor thus encodes the collective
orientational arrangement of the atomic sites. By construction, $S_g=1$ when all H$_2$ molecular axes are perfectly aligned along the crystallographic $c$ axis, while $S_g=0$ corresponds to a completely random, isotropic distribution of the molecular orientations in the simulation cell.

We then apply a \emph{quantum dressing} to this geometric factor, obtained from the probability density of the quantum embedded model at pressure P and temperature T. For this purpose, we define the alignment tensor
\begin{equation}
\mathbf{Q}_q = \frac{3}{2}\int 
\rho(\mathbf{r})\,
\mathbf{u}_{\mathbf{r}}\mathbf{u}_{\mathbf{r}}\,\mathrm{d}^{3}\mathbf{r}
-\frac{1}{2}\mathbf{I},
\label{Sq}
\end{equation}
where $\mathbf{u}_{\mathbf{r}}$ is the unit vector along $\mathbf{r}$,
and $\rho(\mathbf{r})$ is the diagonal part of the thermal density matrix. We define the quantum contribution $S_q$ as the largest eigenvalue 
of $\mathbf{Q}_q$.
By construction, $S_q=1$ when 
the quantum distribution of the
H$_2$ molecular axes is perfectly aligned 
along a given direction while $S_q=0$ corresponds to an isotropic quantum distribution.
This quantum dressing ensures that the final orientation factor incorporates both the orientation linked to the underlying geometry ($S_g$) and the intrinsic thermal quantum fluctuations encoded in the density matrix
($S_q$), in the same way as diffraction separates lattice geometry from the internal structure of the scatterer.

With this definition, the orientation factor $S(P,T)$ can be mapped across the phase diagram, as shown in Fig.~\ref{fig:PhaseDiagram}.

$S(P,T)$ can also be computed from PIMD simulations at pressure $P$ and temperature $T$, whenever available, instead of the quantum embedded model. In this case, the orientation factor reads
$S_{\mathrm{PIMD}} = \int dX \,\rho(X)\, \frac{1}{N N_b}\sum_{i=1}^{N N_b}\left(\tfrac{3}{2}\cos^2\theta_i(X)-\tfrac{1}{2}\right)$,
where the integral runs over the path configurations $X$, and the sum runs over all $N$ H$_2$ molecules and the beads. Here, $\rho(X)$ denotes the normalized configurational probability density sampled during the path-integral molecular dynamics,
so that the integral represents an ensemble average over beads and time steps. Notice how the latter equation contains implicitly both a geometrical contribution, given by the orientational arrangement of the H$_2$ molecules in the supercell, and thermal-quantum fluctuations, provided by the path-integral formalism. The orientation factor obtained from the PIMD simulations $S_{\mathrm{PIMD}}$ at 300K is plotted in Fig.~\ref{fig:final}, by neglecting spin isomers effects and relying on the similar density distribution of the ${\mathcal J}=0$ and ${\mathcal J}=1$ states at room temperature.

To locate the transition between the quantum plastic and orientationally ordered phases, we employ a complementary global order parameter $M_{\mathrm{tot}}$. 
It is defined as a structure–factor measure of orientational coherence, constructed from angular phases chosen such that symmetry–related orientations (phase/antiphase configurations, e.g. in the herringbone pattern) interfere constructively. Randomly oriented H$_2$ molecules then cancel out ($M_{\text{tot}}\approx0$), while herringbone or nematic H$_2$ molecules add coherently ($M_{\text{tot}}\approx1$). The precise mathematical definition, including the angular cumulative distribution function transforms and rank-m order parameters, is given in Section~\ref{sec:orderparam} of the SI.
In molecular dynamics trajectories sampled over the $(P,T)$ grid, orientational transitions are identified by the vanishing of $M_{\mathrm{tot}}$, which signals the loss of orientational order in the H$_2$ sublattice. For each pressure, the temperature at which $M_{\mathrm{tot}}$ vanishes is referenced as the transition point.

Since $M_{\mathrm{tot}}$ cannot be evaluated within the quantum embedded model, which describes individually embedded H$_2$, we instead compute the global order parameter from classical MD simulations.
At the classical transition points, we evaluate the corresponding orientation factors, where this time thermal quantum fluctuations in Eq.~\ref{Sq} are replaced by a thermal average performed over the classical canonical ensemble in the simulation cell.
Within the approximation of dominant electrostatic effects, the resulting ensemble-averaged value of $S$ at the classical transition points
provides a threshold, $S^{\mathrm{th}}$, for the critical value of $S$ computed with the quantum embedded model.
Within the embedded framework, the transition temperature $T^*_P$ at pressure $P$ is then obtained from the 
condition
\[
S(P,T^*_P) = S^{\mathrm{th}}
\]
Plotting $T^*_P$ as a function of pressure gives the points of the white dashed transition line shown in Fig.\ref{fig:PhaseDiagram}.
The error bars on $T^*_P$ are obtained by propagating the uncertainty in $S^{th}$: we evaluate the inverse relation at the two bounds $S = S^{th} \pm \sigma_S^{th}$. The temperature uncertainty is then defined as $\Delta T = \tfrac{1}{2}|T_{\rm high}^* - T_{\rm low}^*|$, where $T_{\rm high}^*$ ($T_{\rm low}^*$) corresponds to the temperature extracted from the inverse map at the $S^{th} + \sigma_S^{th}$ ($S^{th} - \sigma_S^{th}$) bound (see Fig.~\ref{fig:Sq} of SI).

Finally, because $S_g$ is computed with respect to the $c$ axis in Eq.~\ref{Sg}, the orientation factor also provides direct information on the nematic orientation of the system, the largest values of $S$ being taken in the nematic aligned phase.

The phonon spectrum and Raman cross section of the system were computed using density functional perturbation theory (DFPT) as implemented in the Quantum ESPRESSO package \cite{RevModPhys.73.515, PhysRevLett.90.036401}, employing a unit cell and a 4×4×4 k-point grid. This is the only calculation performed entirely within a classical nuclear framework, as no implementation currently exists for computing Raman cross sections from PIMD trajectories.

To investigate the relationship between H$_2$ orientational order, H$_2$O–H$_2$ interactions, and proton symmetrization within the water sublattice, the radial distribution functions $g(r)$ as a function of pressure were computed from our MD trajectories for H$_\mathrm{w}$–O, H$_\mathrm{h}$–O, and O–O pairs. Here, H$_\mathrm{w}$ denotes protons belonging to the water sublattice, H$_\mathrm{h}$ refers to protons from H$_2$ molecules, and O represents oxygen atoms.

\section*{Acknowledgments}
We acknowledge the European Synchrotron Radiation Facility for provision of synchrotron radiation facilities at the ID15B and ID27 beam lines and assistance from G. Garbarino and S. Gallego Parra.
We acknowledge U. Ranieri, A. Falenti, W. Kuhs and L. Ulivi for samples preparation.
L.E.B. and M.R. acknowledge the financial support by the European Union - NextGenerationEU (PRIN N. 2022NRBLPT), the ANR-23-CE30-0034 EXOTIC-ICE. 
L.E.B., T.P. and A.N. acknowledge the financial support by the Swiss National Fund (FNS) under Grant No. 212889. 
S.D.C. acknowledges computational resources from CINECA, proj. IsC90-HTS-TECH and IsC99-ACME-C, and the Vienna Scientific Cluster, proj. 71754 "TEST".
L.R. and A.M.S acknowledge GENCI for providing computational resources on the IDRIS Jean-Zay supercomputing clusters under project numbers 2024-A0160901387, 2025-A0180901387.
L.R. and M.C. acknowledge GENCI for providing computational resources on the IDRIS Jean-Zay supercomputing clusters under project number A0170906493.
M. C. thanks the European High Performance Computing Joint Undertaking (JU) for the partial support through the "EU-Japan Alliance in HPC" HANAMI project (Hpc AlliaNce for Applications and supercoMputing Innovation: the Europe - Japan collaboration).

\section*{Author Contributions}
L.E.B. designed the research. S.D.C.  initiated the theoretical investigation. L.R. and S.D.C. performed the MD simulations and data analysis with guidance from M.C., A.M.S., and L.E.B.. 
T.P., L.A., A.N., R.G., and L.E.B. performed the XRD experiments. T.P., A.N., and M.R. performed the Raman experiments. T.P. and L.A. performed the XRD data analysis, A.N., R.G. and M.R. performed the Raman data analysis, with guidance from L.E.B.. 
L.R., T.P., and A.N. prepared the figures, with inputs from L.E.B., M.C., and A.M.S..
L.E.B, L.R., and T.P. wrote the manuscript with input from M.C., and A.M.S. 
All authors discussed the results and the manuscript.

\bibliography{\bib}
\newpage
\appendix

\section*{Computational supplementary}
\section{Quantum solution of the embedded H$_2$ molecule in the crystal field}
\label{sec:schro}

Building upon the method described in the Supplementary Material of Ref.\cite{PhysRevLett.133.236101}, we evaluated the full, non-separable external potential $V_{\text{ext}}(r, \theta, \phi)$ and solved the corresponding Schrödinger equation for the quantum embedded H$_2$ molecule where $r$ is the H$_2$ molecule elongation, $\theta$ is the polar and $\phi$ the azimuthal angles describing the molecule orientation (see Fig. \ref{fig:distLPLT}).

Quantum embedding leads to a system of quantum rigid rotor in an effective potential representing an isolated H$_2$ in the C2 phase. The associated Schrödinger equation reads:

$$
\left( -\dfrac{\nabla^2}{2\mu} + V_{\text{ext}} \right) \psi(\vec{r}) = \epsilon \psi(\vec{r}),
$$

where $\mu = \dfrac{m_H}{2}$ is the reduced mass, with $m_H$ the mass of a hydrogen atom. Since $V_{\text{ext}} \neq 0$, the energy levels deviate from those of a free rotor, given by $E_l = \dfrac{\hbar^2}{2I} l(l+1)$, where $I = \mu R^2$ is the moment of inertia and $l$ is the angular momentum quantum number.

\begin{figure}[h!]
    \centering
    \includegraphics[width=\textwidth]{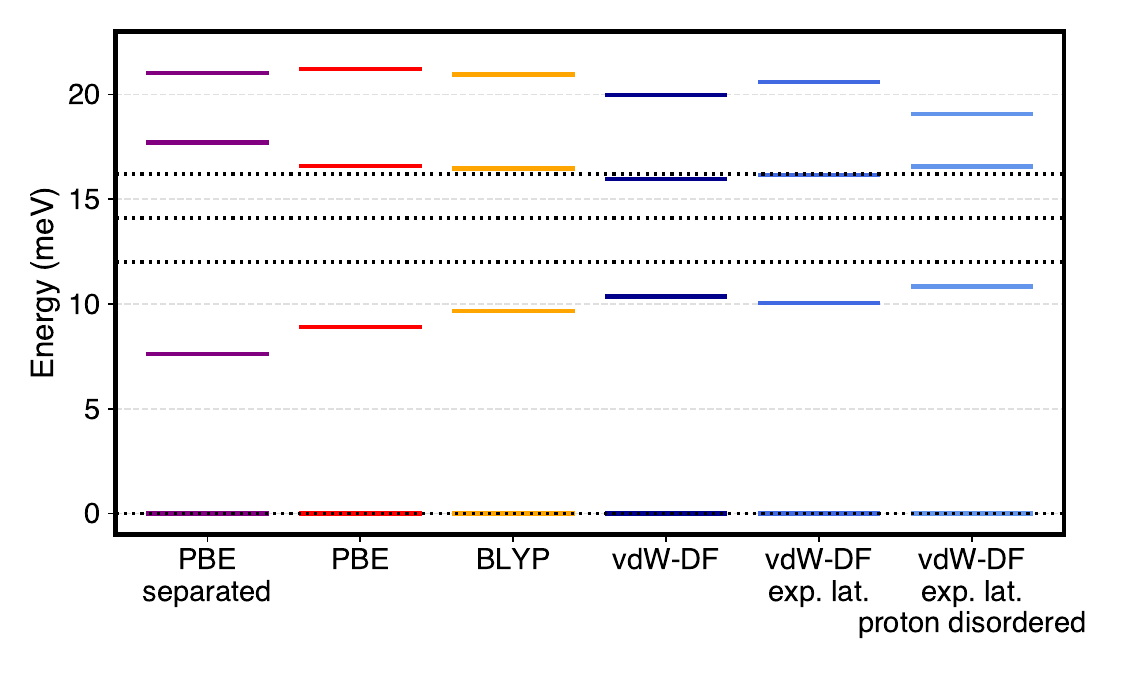}
    \caption{Comparison of the computed eigenvalues at P=3GPa using our method (with various exchange-correlation functionals: PBE, BLYP, and vdW-DF) is shown alongside results obtained with the approach presented in Ref.\cite{PhysRevLett.133.236101} (PBE with a separated potential). The labels vdW exp. lat. and vdW exp. lat. proton disordered refer to calculations using the vdW-DF functional on the C2 structure with experimental lattice parameters, with and without averaging over disordered proton configurations, respectively. Experimental energy levels from Ref.\cite{PhysRevLett.133.236101} are shown as black dotted lines.}
    \label{fig:quantumlevels}
\end{figure}

The external potential takes the form $V_{\text{ext}} = V_{\text{DFT}}(\vec{r}) = V(r, \theta, \phi) = R(r) + P_r(\theta, \phi)$. 
The potential 
has been
computed numerically from the DFT-relaxed structure of the C2 hydrogen hydrate, using several DFT functionals reported in Fig.~\ref{fig:quantumlevels}. A script generated all configurations on a spherical coordinate grid with a fixed center of mass, varying the bond length $r$ and orientation angles $\theta$ and $\phi$. The grid used has dimensions $n_r \times n_\theta \times n_\phi = 64 \times 32 \times 32$.

As in Ref.\cite{PhysRevLett.133.236101}, a self-consistent DFT calculation was performed in the primitive unit cell for each configuration. Since the simulations are carried out at 0K, the resulting total energies represent the potential experienced by the H$_2$ molecule around the classical equilibrium geometry.
These calculations were performed with Quantum Espresso \cite{Giannozzi_2009, 10.1063/5.0005082}, on a k grid of 5x4x5 and a
plane-waves expansion cut at 300 eV. 

The radial part of the potential $R(r)$ is fitted using a Morse function, while the angular component is interpolated linearly between the computed data points. After having determined numerically the effective potential $V_{\text{ext}}$, we solved the corresponding Schrödinger equation numerically using the Lanczos algorithm in shift-invert mode, following the approach of Ref.\cite{PhysRevLett.133.236101}. At variance with Ref.\cite{PhysRevLett.133.236101}, in the present case we took into account the full 3D potential in a non-separable form, by developing the Hamiltonian matrix elements on a basis set of the type $\psi_n^{lm}(r, \theta, \phi) = \chi_n(r) Y_l^m(\theta, \phi)$.  Here, $\chi_n(r)$ are the eigenfunctions of the Morse potential obtained by solving the one-dimensional Schrödinger equation for the fitted radial part, and $Y_l^m(\theta, \phi)$ are spherical harmonics, which form a complete basis set for the angular part of a free quantum rotor.

Since the Hamiltonian can be decomposed as $\hat{H} = \hat{T} + \hat{V}$, with:

$$
\hat{T} = \frac{\hbar^2}{2\mu} \left( \frac{1}{r^2} \frac{\partial}{\partial r} \left( r^2 \frac{\partial}{\partial r} \right) + \frac{1}{r^2} \hat{L}^2 \right), \quad \text{and} \quad \hat{V} = R(r) + P_r(\theta, \phi),
$$

the Hamiltonian matrix elements in this basis are given by the following terms:

\begin{enumerate}
    \item Kinetic term: $
\langle n' l' m' | \hat{T} | n l m \rangle = \frac{\hbar^2}{2\mu} \left[ \int_0^\infty \frac{d\chi^*_{n'}(r)}{dr} \frac{d\chi_n(r)}{dr} r^2\, dr - \int_0^\infty \chi^*_{n'}(r) \chi_n(r)\, l(l+1)\, dr \right] \delta_{ll'} \delta_{mm'}.
$
    \item Potential term: $
\langle n' l' m' | \hat{V} | n l m \rangle = \int_0^\infty \chi^*_{n'}(r) R(r) \chi_n(r) r^2\, dr\, \delta_{ll'} \delta_{mm'}
+ \int_0^\infty \chi^*_{n'}(r) \chi_n(r) r^2\, dr \cdot A_{l' m' l m}(r),
$ where the angular coupling coefficient is defined as:\\ 
$ A_{l' m' l m}(r) = \int d\Omega\, Y_{l'}^{m'*}(\theta, \phi)\, P_r(\theta, \phi)\, Y_l^m(\theta, \phi).
$

\end{enumerate}

All matrix elements can be computed numerically based on the evaluated potential and the chosen basis. The angular integrals $A_{l' m' l m}(r)$ are evaluated on the $r$-grid and interpolated accordingly.

Our method is compared to that of Ref. \cite{PhysRevLett.133.236101} in Fig. \ref{fig:quantumlevels}. When the full, non-separable potential is taken into account, the energy level splitting is reduced compared to the separated approach (PBE separated vs. PBE), where the radial and angular parts are integrated independently. This difference is due to a more accurate treatment of the interaction potential in the present formalism.

We also investigated the impact of the exchange-correlation functional within our method. The use of vdW-DF leads to results that are in better agreement with experiment than those obtained with PBE for DFT energy estimation, consistent with previous findings \cite{Clay_2014}. Since vdW-DF tends to slightly overestimate the lattice parameters, we computed the eigenvalues using the experimental volume at 3 GPa. As expected, a slight increase in the splitting is observed compared to the standard vdW-DF calculation. This is consistent with the pressure dependence of the splitting, as shown in Fig. \ref{fig:simulation sum} (c).

Next, we evaluated the influence of proton disorder on the eigenvalues. To do this, we generated external potentials for a set of proton-disordered configurations that satisfy the ice rules. These potentials exhibit significant variations in magnitude. For each configuration, the corresponding eigenvalues were computed and the average is obtained using an arithmetic mean. As shown in Fig. \ref{fig:quantumlevels}, this leads to a substantial reduction in the energy level splitting, bringing it closer to the experimental values. This reduction is attributed to the decrease in anisotropy of the effective potential when proton disorder is included.

Incorporating all these effects—namely, the use of vdW-DF and averaging over multiple proton-disordered structures—significantly increases the computational cost. In this work, our aim is to provide insight into the orientational transitions and the coupling between the H$_2$ and H$_2$O sublattices, as well as to highlight the role of quantum effects. Furthermore, since both the \emph{ab initio} molecular dynamics (AIMD)  and path-integral molecular dynamics simulations were carried out using the PBE functional due to computational cost, all quantum embedded calculations shown in the main paper have been performed with the same functional and in the proton-ordered configuration (i.e., the ice VIII sub-lattice) for consistency and direct comparison.

\section{Order parameter}
\label{sec:orderparam}
\subsection{Symmetries}

\begin{figure}%
    \centering
    \raisebox{0.3\height}{\subfloat(\centering a){{\includegraphics[width=4cm]{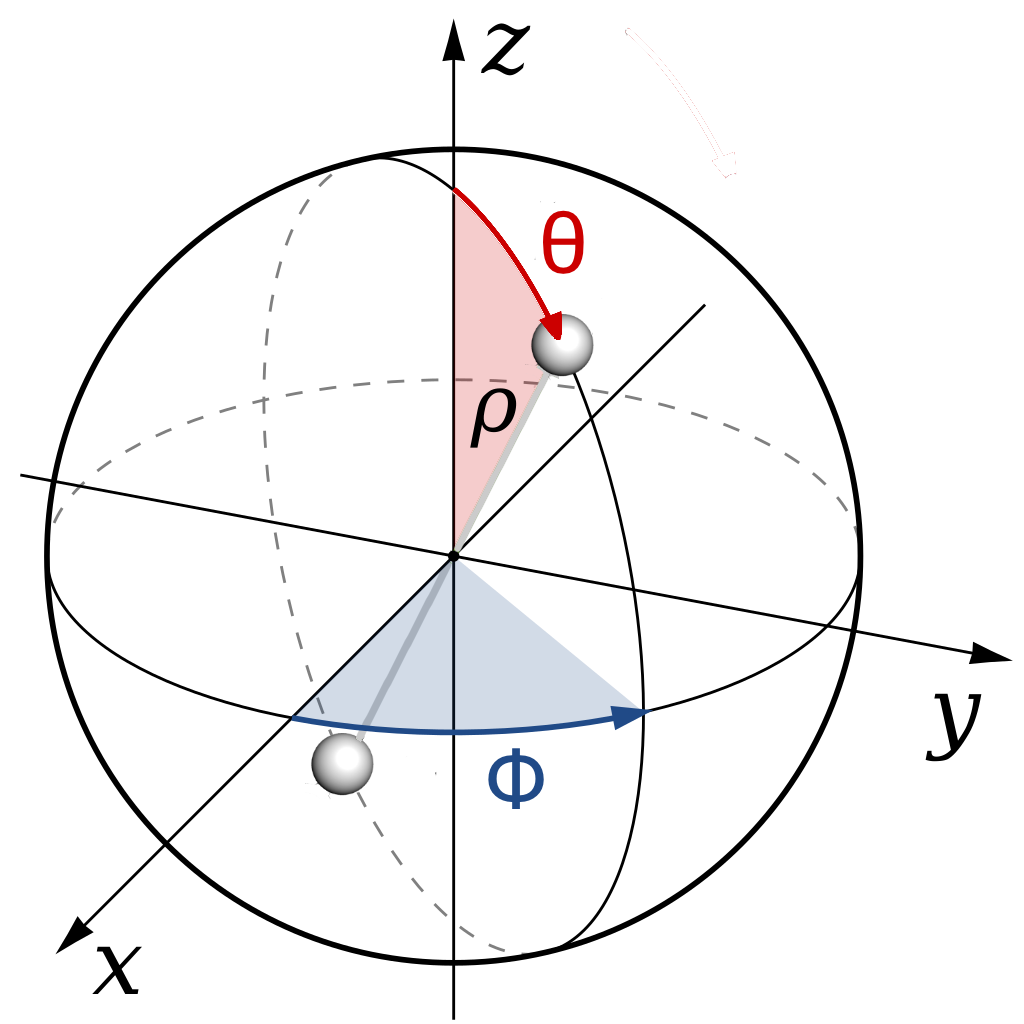} }}}%
    \qquad  
    \subfloat(\centering b){{\includegraphics[width=6cm]{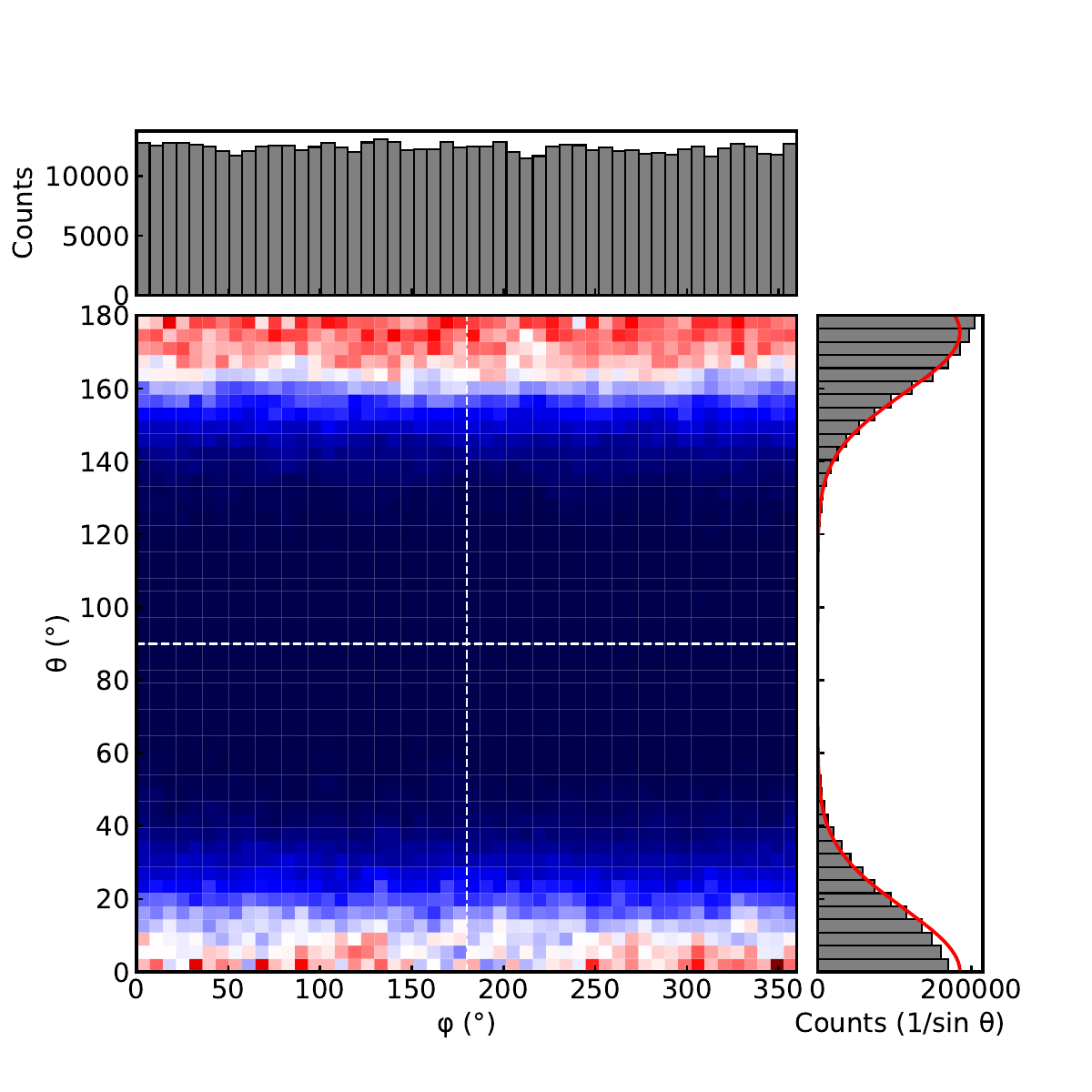} }}%
    \qquad
    \subfloat(\centering c){{\includegraphics[width=6cm]{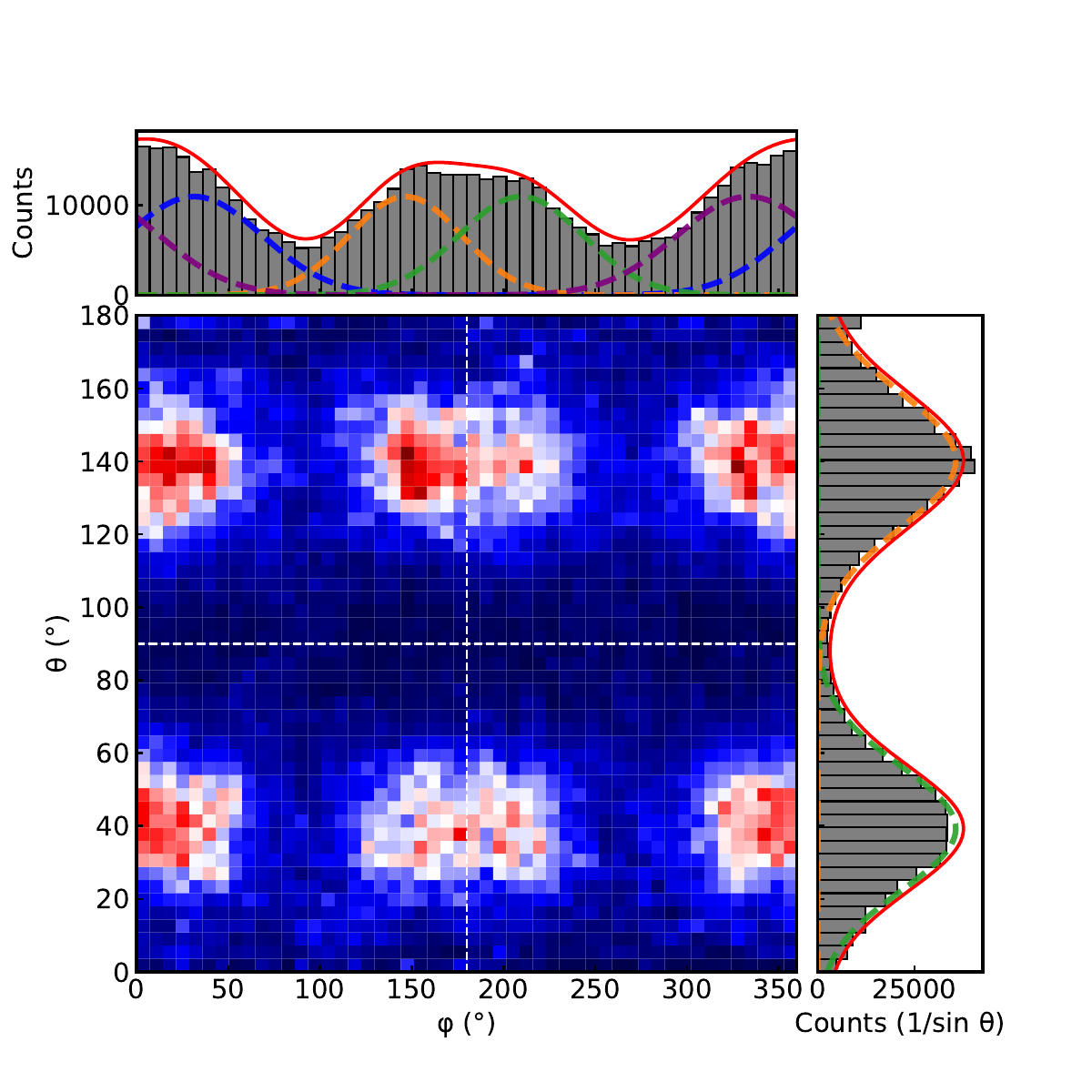} }}%
    \subfloat(\centering d){{\includegraphics[width=6cm]{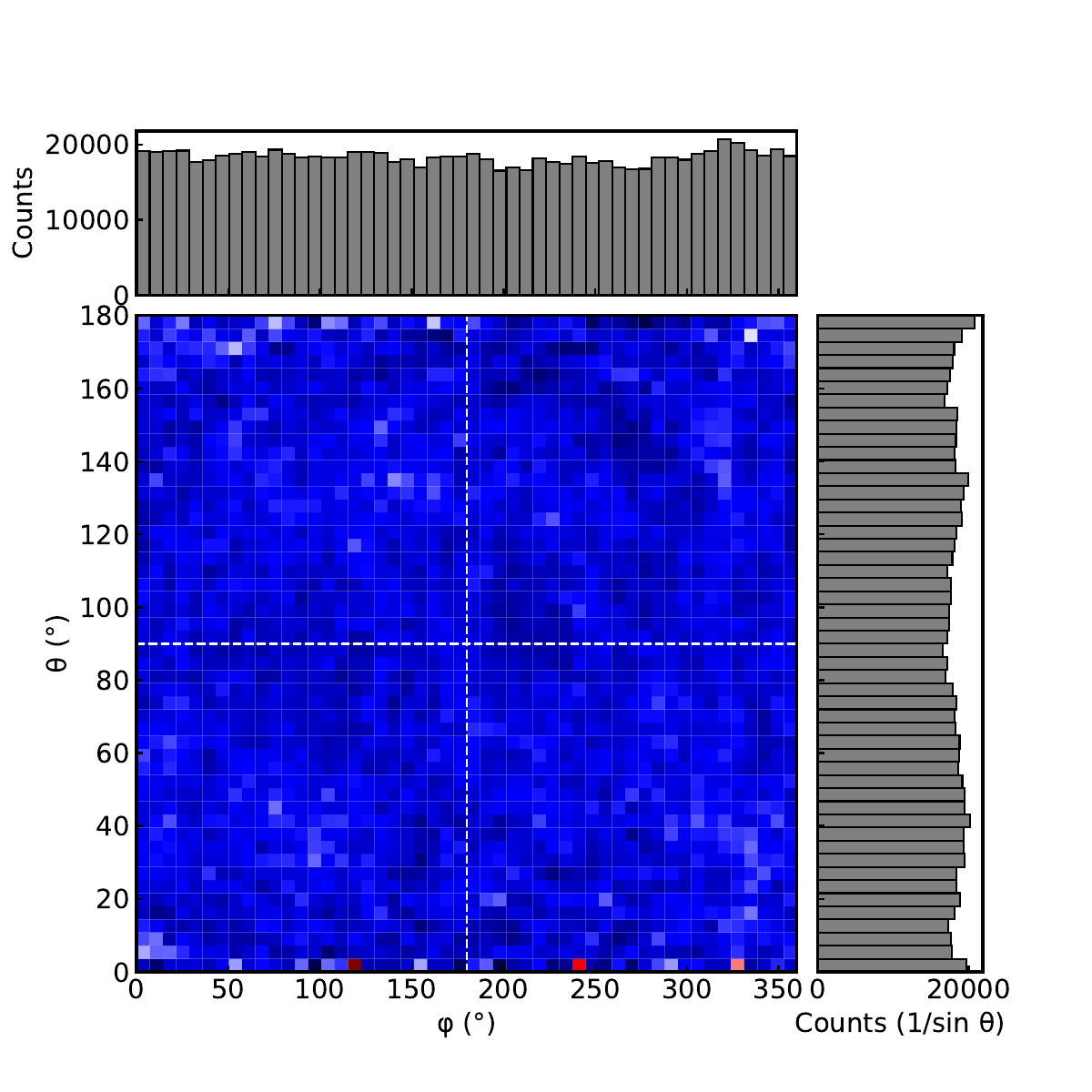} }}%
    \caption{(a) Definition of the spherical coordinates used to describe the orientation of the H$_2$ molecule; the $x$, $y$, and $z$ axes are aligned with the crystallographic axes.
Histogram of the orientation angles sampled by H$_2$ molecules during an ab initio molecular dynamics simulation at $T = 100$ K and $P = 50$ GPa (b), $T = 50$ K and $P = 20$ GPa (c) and $T = 300$ K and $P = 3$ GPa (d).}%
    \label{fig:distLPLT}%
\end{figure}

The orientational order of the \(\mathrm{H_2}\) sub-lattice is monitored 
through a global order parameter.
Figure~\ref{fig:distLPLT} (c) displays a representative two–dimensional
histogram of the polar angle \(\theta\) and the azimuthal
angle \(\phi\) gathered along a
\(T = 50\;\mathrm{K}\), \(P = 20\;\mathrm{GPa}\) molecular-dynamics
trajectory.
Eight bright spots appear, revealing the eight
symmetry orientations available to an
\(\mathrm{H_2}\) molecule in the herringbone phase.
As discussed in the main text, increasing pressure forces the sticks, representing the H$_2$ geometry, to align; in the fully ordered regime, i.e. nematic, the eight spots collapse into two. Since $\phi$ is not well defined for $\theta=0$ we see in practice two bands as shown in \ref{fig:distLPLT} (b).

Although eight maxima are visible, only two
\emph{physically distinct} orientations remain once crystalline
symmetries and the head-to-tail indistinguishability of the sticks are
taken into account:

\begin{enumerate}[label=(\roman*)]
  \item \textbf{Polar symmetry.}  
        Because the colatitude of an unoriented stick is defined
        modulo inversion, the two spots at
        \(\theta_1\) and \(\theta_2=\pi-\theta_1\) correspond to the
        \emph{same} configuration.
  \item \textbf{Azimuthal symmetry.}
        In the longitudinal direction one has
        \(\phi_2=\phi_1+\pi\); the pair
        \((\phi_1,\phi_2)\) therefore describes a single molecular
        orientation.
        The remaining spots at
        \(\phi_3=\pi-\phi_1\) and
        \(\phi_4=2\pi-\phi_1\) are likewise equivalent to each other.
\end{enumerate}

Accordingly, at any pressure the eight histogram maxima partition into
two families of four spots, leaving only \emph{two} genuinely different
orientational states in the herringbone phase. The quantum plastic case, with no orientational order, is shown in \ref{fig:distLPLT} (d).

\subsection{Global order parameter}

As explained in the Methods section, we define the order parameter to locate the transition between the quantum plastic and orientationally ordered phases. This parameter is designed to capture the orientational coherence of the H$_2$ sublattice. It is constructed as a structure factor, which vanishes when the H$_2$ molecules are randomly oriented in the quantum plastic phase, and interferes constructively when the molecules are in the herringbone or nematic phase.

The global order parameter, \(M_{\text{tot}}\), can be separated into two components: \(M_\theta\), which characterizes the order in \(\theta\), and \(M_\phi\), which characterizes the order in \(\phi\):
\[
  M_{\text{tot}} = \sqrt{M_\theta^2 + M_\phi^2}.
\]

These components can be written individually as:
\[
  M_\theta = \left\langle \left|\psi_\theta(t)\right| \right\rangle_t,
  \qquad
  M_\phi = \left\langle \left|\psi_\phi(t)\right| \right\rangle_t.
\]

Here, \(\psi_\theta(t)\) and \(\psi_\phi(t)\) are the rank-\(m\) order parameters with \(m_\theta = 4\) and \(m_\phi = 2\) at time \(t\). They are defined as:
\[
  \psi_\theta(t) = \frac{1}{N} \sum_{\mu=1}^{N} e^{i m_\theta \theta_\mu^{\text{corr}}(t)},
  \qquad
  \psi_\phi(t) = \frac{1}{N} \sum_{\mu=1}^{N} e^{i m_\phi \phi_\mu^{\text{corr}}(t)},
\]

The sum runs over the \(N\) H$_2$ molecules in the system (\(\mu = 1, \dots, N\)). The angles \(\theta_\mu^{\text{corr}}\) and \(\phi_\mu^{\text{corr}}\) are the polar and azimuthal angles of the H$_2$ molecules, corrected to give the same phase in symmetrically equivalent configurations, as discussed in the previous section. The goal is to have an order parameter that vanishes when the molecules are disordered and equals 1 when the molecules are perfectly aligned, following the structure factor concept.

For each molecule, the bond vector \(\mathbf{r}_\mu\) is expressed in spherical coordinates \((r_\mu, \theta_\mu, \phi_\mu)\) in the crystal frame:
\[
  \theta_\mu = \arccos\left( \frac{y_\mu}{r_\mu} \right),
  \qquad
  \phi_\mu = \arctan2(z_\mu, x_\mu) \in [-\pi, \pi[.
\]

In order for the symmetric configurations mentioned above to yield the same phase in the herringbone-like situation, one possible estimator for the \(\theta\) angle is \(\sin\theta\). We flatten the distribution of \(\sin\theta\) via the cumulative distribution function (CDF) transform:
$$
\boxed{\theta_\mu^{\text{corr}} = \frac{\pi}{2} \left( 1 - \cos\theta_\mu \right) \in [0, \pi[}.
$$

This transformation ensures that, for an isotropic rotor, the phase \(\exp(i \theta_\mu^{\text{corr}})\) is uniformly distributed on the circle, allowing the terms to cancel out statistically.

A similar approach is applied to \(\phi\), since symmetric configurations have the same phase when using \(|\sin\phi|\). Then, using the CDF to obtain a continuous uniform distribution, we have:
$$
\boxed{\phi_\mu^{\text{corr}} = 2 \arcsin\left( |\sin\phi_\mu| \right) \in [0, 2\pi[}.
$$

\paragraph{Baseline}
To account for the finite size of our simulation (both in time and length), we subtract a baseline: the same order parameter computed for a system of identical size and the same number of time steps, but with completely random H$_2$ positions.\\

\begin{figure}%
    \centering
    \includegraphics[width=10cm]{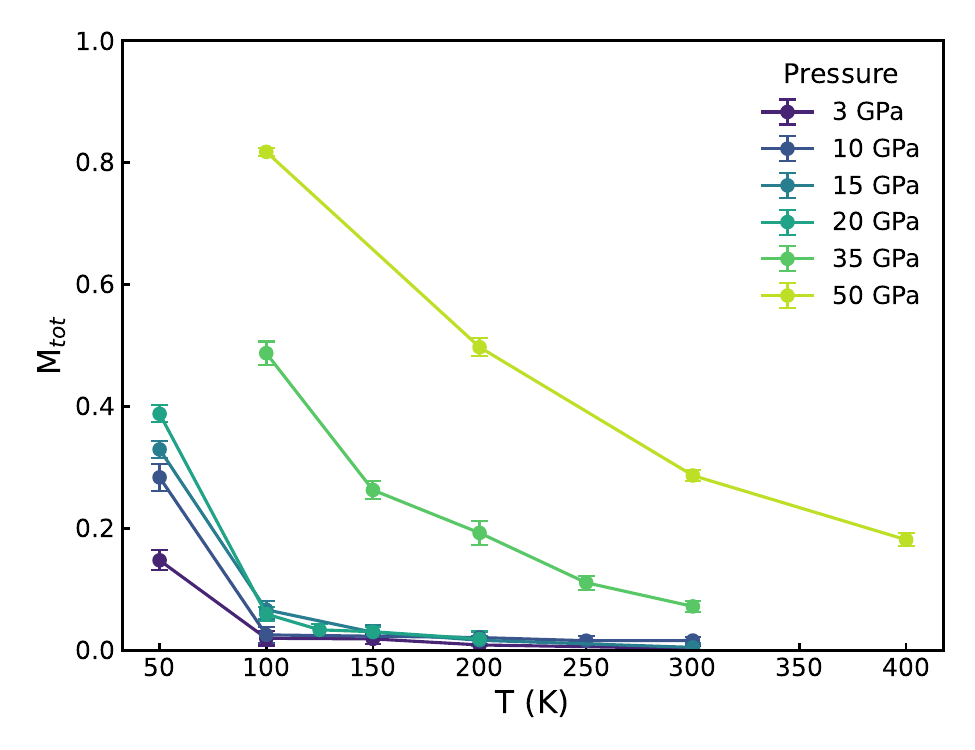} %
    \caption{Global order parameter, \(M_{\text{tot}}\), plotted as a function of temperature for AIMD trajectories.}%
    \label{fig:orderparam}%
\end{figure}

We show in Fig. \ref{fig:orderparam} the evolution of the global order parameter as a function of temperature for different pressure values. For example, at 3 GPa, we observe that \(M_{\text{tot}}\) vanishes at 100 K. 

Since \(M_{\text{tot}}\) is a global parameter, it cannot be evaluated within the framework of the quantum embedded model, as it describes a single embedded H$_2$. As shown in the Methods section, in the quantum embedded model, the appropriate parameter is the orientation factor \(S\). Thus, to predict the transition line between the disordered and ordered phases in the quantum embedded model, we need to find a suitable criterion, i.e., the critical value of \(S\), that can indicate the transition from quantum plastic to ordered phases.

To do this, we compute the orientation factors \(S\) at the transition points predicted by \(M_{\text{tot}}\) from molecular dynamics simulations at different pressures. This provides a criterion to locate the phase transition in terms of \(S\) in the quantum embedded framework. 

\begin{figure}%
    \centering
    \includegraphics[width=10cm]{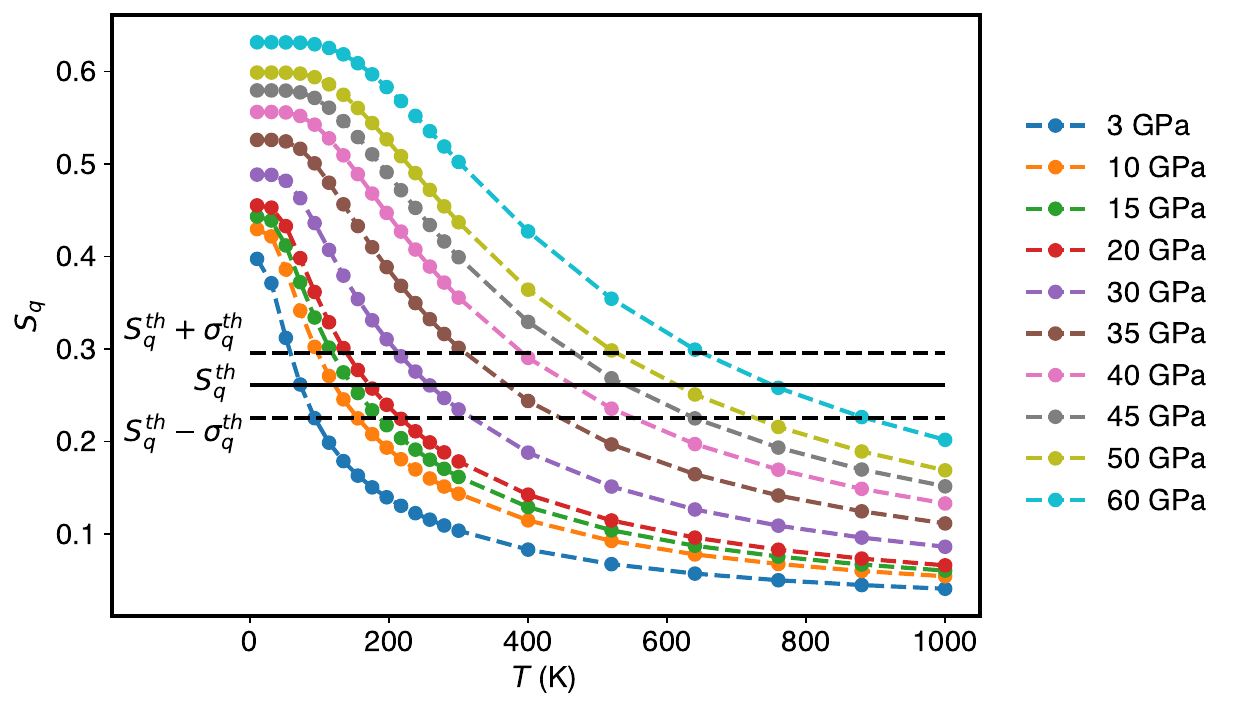} %
    \caption{Quantum contribution \(S_q\) to the orientation factor evaluated from the quantum embedded model as a function of temperature for different pressures, together with the computed threshold $S_q^{th}$ used to determine the transition temperature. In dashed lines are displayed the upper and lower bounds $S_q^{th} \pm \sigma_q^{th}$ used to determine the error bars on the transition temperature.}%
    \label{fig:Sq}%
\end{figure}

In practice, to evaluate \(S\) from molecular dynamics runs at the transition point (for example, at 100 K and 3 GPa, as shown in Fig. \ref{fig:orderparam}), we need an estimate of \(S_q\) as defined in Eq.~\ref{Sq} in the main paper. In the quantum embedded model, \(S_q\) is defined using the alignment tensor involving the thermal density matrix, which is not obtained in the classical molecular dynamics (MD) runs. Under the approximation of dominant charge distribution effects via electrostatic interactions, we can replace thermal quantum effects by classical thermal fluctuations, and estimate the critical value of \(S_q\) for the quantum model as a thermal average performed over the classical canonical ensemble in the simulation cell at the classical transition point. 
Since the effective potential $V_{\text{ext}}$ of the quantum embedded model is obtained from the classical geometry at zero temperature, 
the geometric contribution \(S_g\) to the orientation factor is the same 
for both quantum embedded and classical MD frameworks.
Thus, finding the threshold for \(S_q\) is sufficient. Moreover, we found that within the pressure and temperature resolution of our classical MD simulations, \(S_q\) at the transition can be safely taken as pressure independent. By averaging the obtained estimations of \(S_q\) at the transition point over pressure (see Tab. \ref{tab:Sq}), we determined the threshold \(S_{q}^{\text{th}} = 0.26 \pm 0.04\) for the fluctuating part. Applying this to the datapoints plotted in Fig. \ref{fig:Sq}, we can find the temperatures at which the transition occurs in the quantum embedded framework. We also displayed the upper and lower bounds $S_{q}^{\text{th}} \pm \sigma_q^{th}$ used to determine the errors on temperature in Fig. \ref{fig:Sq}, as explained in Methods.

\begin{table}[]
\centering
\begin{tabular}{l|cccc}
\textbf{P (GPa)} & \textbf{3} & \textbf{10} & \textbf{15} & \textbf{20} \\ \hline
T (K)            & 100        & 100         & 125         & 150         \\ \hline
$S_q$            & $0.24 \pm 0.08$       & $0.25 \pm 0.08$       & $0.27 \pm 0.07$       & $0.27 \pm 0.07$       \\ \hline
\end{tabular}
\caption{Temperature of transition T estimated within our temperature resolution from Fig \ref{fig:orderparam} and the associated $S_q$ at the transition point.}
\label{tab:Sq}
\end{table}

\section{Free Rotor to Quantum Oscillator}

\begin{figure}
    \centering
    \includegraphics[width=0.9\linewidth]{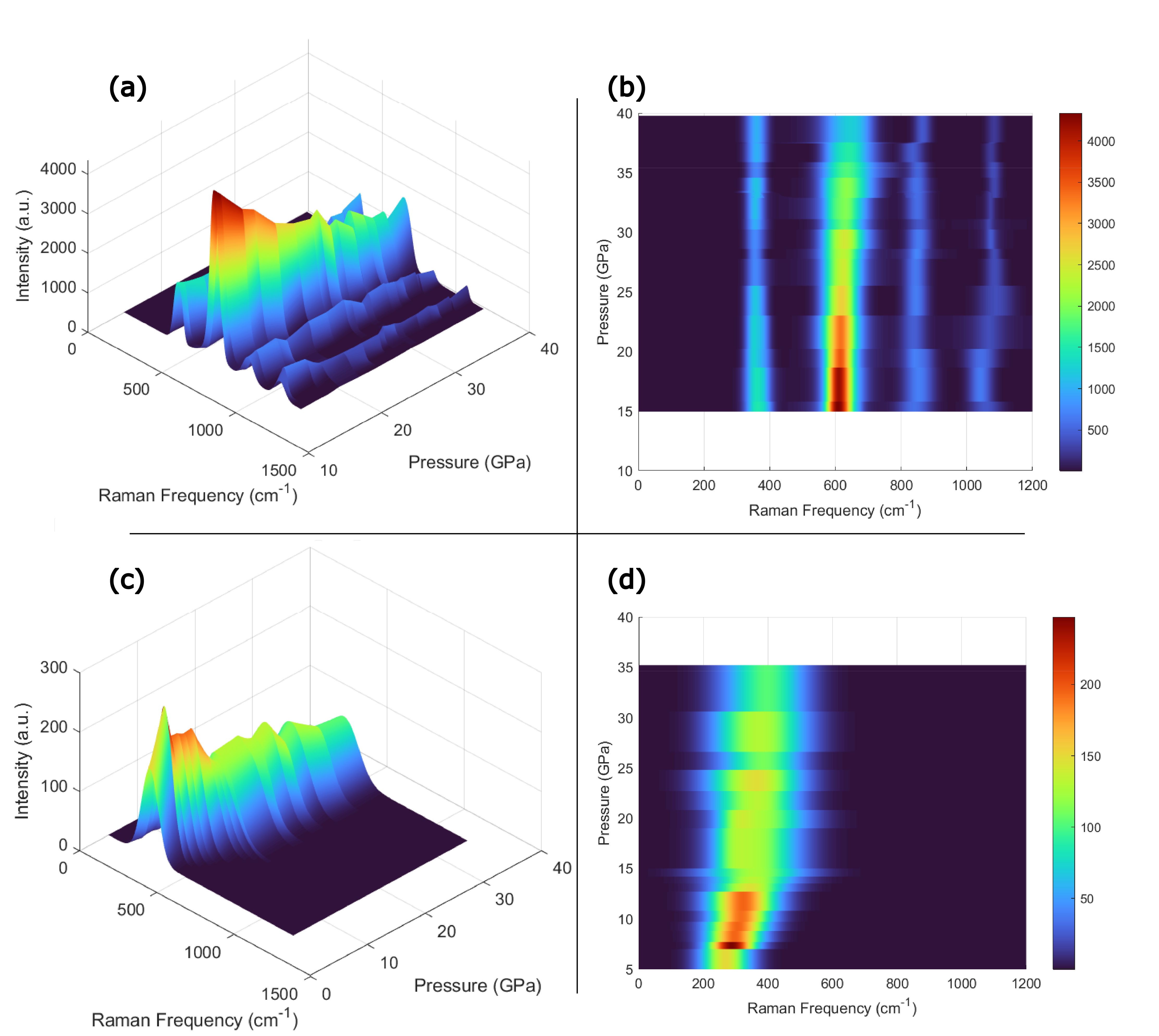}
    \caption{Raman shift for $H_2-H_2O$ and $D_2-D_2O$ as pressure is increased to 40\,GPa at 300\,K, visualised in a 3D-surface plot and a 2D-heat plot. Panels (a) and (b) show, with increasing frequency, the four $S_0(0)$, $S_0(1)$, $S_0(2)$ and $S_0(3)$ rotons of $H_2-H_2O$ in pressure, whilst panels (c) and (d) only show the weak $S_0(1)$ roton of $D_2-D_2O$.}
    \label{fig:3DRaman}
\end{figure}

Figure~\ref{fig:3DRaman} is a visualisation of how the rotons of $H_2-H_2O$ and $D_2-D_2O$ change as the pressure is increased to 40\,GPa at 300\,K. In panels ~\ref{fig:3DRaman}(a) and (b), the $S_0(0)$, $S_0(1)$, $S_0(2)$ and $S_0(3)$ rotons are shown with incresing Raman shift frequency, from left to right, respectively. The $S_0(1)$ is the most intense. The intensity of all the rotons decrease as the pressure is increased, whilst the peaks significantly broaden. 

The evolution of the rotons of $D_2-D_2O$, shown in panels ~\ref{fig:3DRaman}(c) and (d), is different since the Raman spectrometer that was used for the experiments could not well resolve the $S_0(0)$ roton, and they were exceedingly weak in comparison to the $H_2-H_2O$ case as well, as indicated in the intensity scale of the different panels. Nonetheless, the $S_0(1)$ roton was clearly seen and traced in pressure. This peak broadens significantly more than the equivalent HH $S_0(1)$ roton. 

\begin{figure}
    \centering
    \includegraphics[width=0.98\linewidth]{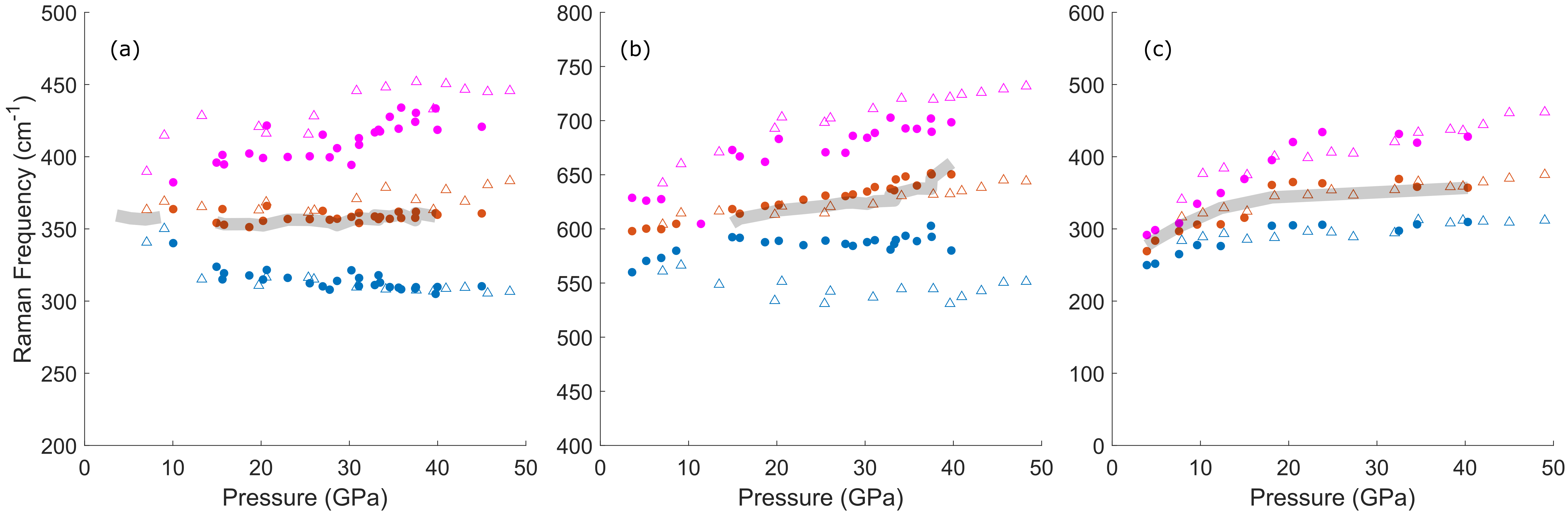}
    \caption{High-Pressure Raman measurements of the rotons of $H_2-H_2O$ and $D_2-D_2O$. The different roton components $S_0(x)_0$, $S_0(x)_1$ and $S_0(x)_2$, are shown by the blue, orange and magenta branches respectively. Open triangles represent measurements of pure $H_2$ or $D_2$ as reported by Pena-Alvarez \textit{et al.} \cite{Pena-Alvarez2020}, filled circles represent measurements of $H_2-H_2O$ or $D_2-D_2O$ up to 40\,GPa, and the thick grey line represents the position of the 'Centre of Mass' of the entire $S_0(x)$ peak. Panel (a) shows the $S_0(0)$ roton for $H_2-H_2O$, panel (b) shows the $S_0(1)$ roton for $H_2-H_2O$ and panel (c) show the $S_0(1)$ roton for $D_2-D_2O$.}
    \label{fig:RamanH2D2}
\end{figure}

Figure~\ref{fig:RamanH2D2} shows the general broadening of the $S_0(0)$ and $S_0(1)$ rotons in $H_2-H_2O$ and $D_2-D_2O$ as pressure is increased to 40\,GPa at 300\,K. The rotons were modelled with three components, referring to the $S_0(x)_0$, $S_0(x)_1$ and $S_0(x)_2$ modes ($x = [0,1]$). These were iteratively fitted, using the previous fitting to ensure that the peaks evolved naturally from pressure to pressure, and that there were no sudden changes to the fitting regime. As can be seen, our measured experimental data has been compared to the the experimental data of pure $H_2$ and $D_2$ as reported by Pena-Alvarez \textit{et al.} \cite{Pena-Alvarez2020}. The rotons of HH shows a good agreement with the rotons of pure $H_2$ or $D_2$. 

Note that the $S_0(1)$ roton progression in pressure is not shown for $D_2-D_2O$. Again, this is due to the fact that the Raman spectrometer that was used did not have a good enough resolution below 180\,$cm^{-1}$ to determine the weak roton from the large background from the elastic line. 

\begin{figure}
    \centering
    \includegraphics[width=0.7\linewidth]{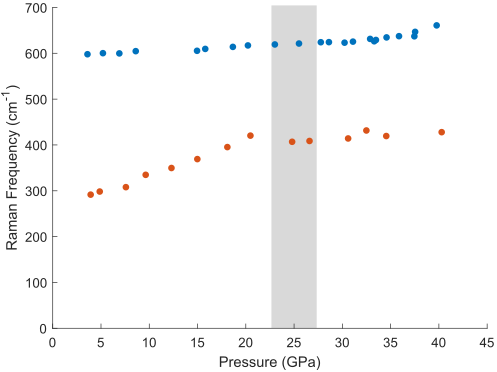}
    \caption{Comparison of the $S_0(1)$ roton 'Centre of Mass' of $H_2-H_2O$ (blue) and $D_2-D_2O$ (orange) in pressure. The grey shaded block indicates the change from a free rotor to a quantum oscillator regime, as detailed in the main text.}
    \label{fig:RamanComp}
\end{figure}

Figure \ref{fig:RamanComp} compares the 'Centre of Mass' of the $S_0(1)$ rotons for $H_2-H_2O$ and $D_2-D_2O$ in pressure. The 'Centre of Mass' of the peak was determined by fitting a single peak to the data, instead of the three components of $S_0(1)_0$, $S_0(1)_1$ and $S_0(1)_2$. The grey band indicates the region where we observe the change in regime from a free roton to a quantum harmonic oscillator potential, and the enclathrated $H_2$ moves from a herringbone-like arrangement and the excitation of the molecules adopt a librational character. 

\end{document}